\documentclass[journal]{IEEEtran}
\usepackage{amsmath,amsfonts}
\usepackage{amsthm}
\usepackage{algorithmic}
\usepackage{algorithm}
\usepackage{array}
\usepackage[caption=false,font=small]{subfig}
\usepackage{textcomp}
\usepackage{stfloats}
\usepackage{xurl}
\usepackage{verbatim}
\usepackage{graphicx}
\usepackage{cite}
\usepackage{xcolor}

\newtheorem{proposition}{Proposition}

\allowdisplaybreaks

\begin{document}

\title{Trajectory PMB Filters for Extended Object Tracking Using Belief Propagation}

\author{Yuxuan Xia, {\'A}ngel F. Garc{\'\i}a-Fern{\'a}ndez, Florian Meyer, Jason L. Williams, Karl Granstr\"{o}m, and Lennart Svensson
\thanks{Y. Xia and L. Svensson are with the Department
of Electrical Engineering, Chalmers University of Technology, Gothenburg, Sweden. E-mail: firstname.lastname@chalmers.se. A. F. Garc{\'\i}a-Fern{\'a}ndez is with the Department of Electrical Engineering and Electronics, University of Liverpool, Liverpool, United Kingdom, and also with the ARIES research centre, Universidad Antonio de Nebrija, Madrid, Spain. F. Meyer is with the Scripps Institution of Oceanography and the Department of Electrical and Computer Engineering, University of California San Diego, La Jolla, CA, USA. J. L. Williams is with the Commonwealth Scientific and Industrial Research Organization, Brisbane, Australia. K. Granstr\"{o}m is with Zoox Inc., San Francisco, CA, USA. The work of K. Granstr\"{o}m was done when he was with Chalmers University of Technology.}
}



\maketitle

\begin{abstract}
In this paper, we propose a Poisson multi-Bernoulli (PMB) filter for extended object tracking (EOT), which directly estimates the set of object trajectories, using belief propagation (BP). The proposed filter propagates a PMB density on the posterior of sets of trajectories through the filtering recursions over time, where the PMB mixture (PMBM) posterior after the update step is approximated as a PMB. The efficient PMB approximation relies on several important theoretical contributions. First, we present a PMBM conjugate prior on the posterior of sets of trajectories for a generalized measurement model, in which each object generates an independent set of measurements. The PMBM density is a conjugate prior in the sense that both the prediction and the update steps preserve the PMBM form of the density. Second, we present a factor graph representation of the joint posterior of the PMBM set of trajectories and association variables for the Poisson spatial measurement model. Importantly, leveraging the PMBM conjugacy and the factor graph formulation enables an elegant treatment on undetected objects via a Poisson point process and efficient inference on sets of trajectories using BP, where the approximate marginal densities in the PMB approximation can be obtained without enumeration of different data association hypotheses. To achieve this, we present a particle-based implementation of the proposed filter, where smoothed trajectory estimates, if desired, can be obtained via single-object particle smoothing methods, and its performance for EOT with ellipsoidal shapes is evaluated in a simulation study.
\end{abstract}

\begin{IEEEkeywords}
Multi-object tracking, extended object tracking, random finite sets, sets of trajectories, factor graph, particle belief propagation.
\end{IEEEkeywords}

\section{Introduction}
Multi-object tracking (MOT) refers to the process of estimating object trajectories of interest based on sequences of noisy sensor measurements obtained from multiple sources \cite{challa2011fundamentals,streit2021analytic}. Conventional MOT algorithms are usually tailored to the point object assumption, where each object is modelled as a point without spatial extent, and it gives rise to at most one measurement at each time step. This assumption is, however, unrealistic for modern high-resolution radar and Lidar sensors, for which it is common that an object gives rise to multiple measurements per time scan. The tracking of such an object is called extended object tracking (EOT), and overviews of EOT literature can be found in \cite{extendedoverview,granstrom2022tutorial}. The focus of this paper is on multiple EOT.

Various extended object measurement models exist in the literature, including, e.g., the Poisson spatial model \cite{ppp}, the set cluster process \cite{swain2010extended,honer2020bayesian}, the Set of Points on a Rigid Body model  \cite{buhren2006simulation,hammarstrand2012extended}, and the physics-based model \cite{knill2016direct}. Among these different measurement models, it is most common to use the Poisson spatial model, where the set of object detections is modelled by an inhomogeneous Poisson point process (PPP). The PPP measurement likelihood has a simple factorization, which avoids explicit associations between measurements and points on objects, thereby making it convenient to use in EOT.

The Poisson spatial model has been used in many multiple EOT algorithms, which can be, in general, summarized into two categories. The first category contains methods based on random vectors, including, e.g., the multiple hypothesis tracker (MHT) \cite{coraluppi2018multiple}, the probabilistic MHT \cite{tang2019seamless}, the joint (integrated) probabilistic data association filter \cite{streit2016jpda,yang2018linear,yang2020marginal}, graph-based filters \cite{florian2020scalable,florian2021scalable,yuansheng2022loopy} using belief propagation (also known as the sum-product algorithm (SPA)), and the box particle filter \cite{de2018box}. The second category contains methods based on random finite sets (RFSs) \cite{rfs}, including the probability hypothesis density (PHD) filter \cite{swain2012phd,phdextended2,phdextended3,bohler2019stochastic,yang2019network,sjudin2021extended}, the cardinalized PHD filter \cite{cphdextended}, filters based on labelled RFSs \cite{lmbextended,hirscher2016multiple,daniyan2018bayesian,scheel2018tracking}, the Poisson multi-Bernoulli mixture (PMBM) filter \cite{pmbmextended2,soextended,xia2019extended,garcia2021poisson}, and its approximation the Poisson multi-Bernoulli (PMB) filter \cite{frohle2020decentralized,xia2021poisson}.

For estimating object trajectories, MOT methods based on random vectors link an object state estimate with a previous estimate or declare the appearance of a new object. For RFSs-based MOT methods, one approach to estimating trajectories is to add a unique label to each single-object state such that each object can be identified over time \cite{yang2019network,lmbextended,hirscher2016multiple,daniyan2018bayesian}. This track labelling procedure may work well in some cases, but it often becomes problematic in challenging scenarios \cite{aoki2016labeling,granstrom2018poisson,garcia2019multiple}. A more advantageous approach is to compute the multi-object posterior on sets of trajectories \cite{garcia2019multiple}, which captures all the information about the trajectories. In the context of EOT, this has led to the development of the trajectory PHD filter \cite{sjudin2021extended} and the trajectory PMBM (TPMBM) filter \cite{xia2019extended}.

Common to all the multiple extended object filters is that the data association problem, due to the unknown correspondences between measurements and objects, needs to be efficiently addressed to keep the computational complexity at a tractable level. A common way is to handle the data association problem in two stages: first, clustering algorithms are used to find a set of reasonable measurement partitions; second, for each partition, the explicit assignment of measurement clusters to objects is either avoided using the principle approximation \cite{phdextended2,phdextended3,yang2019network,cphdextended} or obtained using a 2D assignment algorithm \cite{lmbextended,hirscher2016multiple,daniyan2018bayesian,pmbmextended2,xia2019extended,garcia2021poisson,frohle2020decentralized,xia2021poisson}. These two-step approaches based on clustering and assignment (C\&A) usually work well in scenarios with well-spaced objects, but they may suffer from a decreased performance in scenarios where objects move in close proximity. To improve the tracking performance in such scenarios, \cite{bohler2019stochastic,soextended} use sampling-based methods, which work directly on the multi-object likelihood and solve the data association problem in a single step without using clustering. The simulation results in \cite{soextended} showed that the sampling-based methods outperform methods based on C\&A in scenarios with both distant and close objects. In addition, a method combining hierarchical clustering and Gibbs sampling is presented in \cite{honer2019gibbs} for automotive applications.  

For both C\&A and sampling-based methods, the exhaustive enumeration of association hypotheses is avoided by truncating hypotheses with negligible weights. Thus, their performance might be limited in scenarios with high data association uncertainty. A more scalable data association method for multiple EOT, which avoids explicit enumeration of local association hypotheses, is to directly compute the marginal multi-object posterior, where the uncertainty of unknown data association is marginalized out \cite{yang2018linear,yang2020marginal,meyer2019data,meyer2019scalable,meyer2020scalable,florian2020scalable,florian2021scalable}. The current state-of-the-art multiple EOT algorithm is the SPA filter proposed in \cite{florian2021scalable} using particle belief propagation (BP), which is enabled by a factor graph representation of the joint posterior of the multi-object states and association variables. Simulation results in \cite{florian2021scalable} show that the SPA filter outperforms a PMBM filter that uses C\&A, in terms of both estimation error and runtime. However, the multi-object models employed in \cite{florian2021scalable} are subject to two approximations, as compared to the standard multi-object models \cite{rfs}. First, newborn objects are always detected with probability one. Second, object survival probability is state-independent. 

For the standard multi-object models \cite{rfs} with PPP birth, the posterior density is of the form PMBM \cite{pmbmpoint} without approximation. The PMBM density has a compact representation of global hypotheses with probabilistic object existence in each Bernoulli component and undetected objects represented by a PPP. The PMBM filtering recursions have been established for both point objects \cite{pmbmpoint2} and extended objects \cite{pmbmextended2}, and the resulting PMBM filters have achieved state-of-the-art performance compared to other RFSs-based filters\footnote{An online course on MOT that explains all details regarding PMBM filtering and introduces sets of trajectories is available at \url{https://www.youtube.com/channel/UCa2-fpj6AV8T6JK1uTRuFpw.}}. Moreover, in the PMBM filters, the density of undetected objects is propagated over time via a PPP, whereas in \cite{florian2021scalable} and many other MOT methods using BP \cite{meyer2018message}, the explicit modelling of undetected objects is ignored, and their PPP intensity becomes zero in the filtering density due to the modelling assumption. To provide full trajectory information, the PMBM filtering recursions have been extended to consider the posterior on sets of trajectories for both point objects \cite{granstrom2018poisson} and extended objects \cite{xia2019extended}. Furthermore, the PMBM filtering recursions are recently extended in \cite{garcia2021poisson} to consider a generalized measurement model, in which each object generates an independent set of measurements, but the filtering recursions in \cite{garcia2021poisson} have not yet been further generalized to consider posterior on sets of trajectories.

In this paper, we present the trajectory PMB (TPMB) filters for EOT, where the marginal association probabilities and Bernoulli densities in the update step are jointly obtained using BP. By doing so, we leverage 1) the PMBM filtering recursions on the posterior of sets of trajectories built on the standard extended object models and 2) the efficient data association method in \cite{florian2021scalable} using BP. As a comparison, in our previous work on extended object PMB filtering \cite{xia2021poisson}, the PMB approximation is obtained by marginalizing the data association uncertainties using truncated global hypotheses. Furthermore, compared to previous works \cite{meyer2018message,kropfreiter2016sequential,kropfreiter2019fast} that apply BP only to the data association variables in RFSs-based methods, this paper presents a complete pipeline for applying BP to RFSs-based methods. That is, starting from the closed-form filtering recursion to the factor graph formulation of the joint posterior of sets of objects and association variables, and the message passing equations.

In our preliminary work \cite{xia2019extended}, we presented the PMBM filtering recursions for multiple EOT on posterior of sets of trajectories and an efficient track-oriented PMBM implementation. In addition, we have proposed multiple EOT using BP in \cite{florian2021scalable} using a random vector-based derivation. This paper is a significant extension of \cite{florian2021scalable,xia2019extended}, and it contains the following contributions:
\begin{itemize}
  \item We extend the PMBM conjugacy on posterior of sets of trajectories to consider a generalized measurement model in \cite{garcia2021poisson}, where each object generates an independent set of measurements.
  \item We present a full pipeline for integrating TPMB filtering and BP data association methods. To do so, we present a factor graph representation of the PMBM set of trajectories posterior and association variables, on an augmented trajectory space. Importantly, by leveraging the PMBM filtering recursions, we obtain a simpler and more general derivation of the factor graph formulation, with undetected objects represented by a PPP, as compared to the derivation in \cite[Section 1]{meyer2021scalable}.   
  \item We present the message passing equations, derived using RFSs without the additional modelling assumptions made in \cite{florian2021scalable}, for running BP on the constructed factor graph. We also present particle implementations for two TPMB filters: one estimates the set of alive trajectories and the other estimates the set of all trajectories, which includes both alive and dead trajectories. In addition, we show how smoothed trajectory estimates, if desired, can be further obtained using backward simulation \cite{lindsten2013backward}.
  \item We evaluate the performance of several implementations of TPMBM and TPMB filters using different data association methods for EOT, including C\&A, sampling-based method and BP, with ellipsoidal shapes in a simulation study. The results demonstrate that the TPMB filter using BP has the best trajectory estimation performance.
\end{itemize}

The rest of the paper is organized as follows. In Section II, we introduce the background on multi-object models, sets of trajectories and PMBM. The problem formulation and PMBM filtering recursions are given in Section III. The factor graph formulation and the corresponding equations for running loopy BP are presented in Section IV and Section V, respectively. The particle implementations of the proposed TPMB filters are provided in Section VI. Simulation results are presented in Section VII, and conclusions are drawn in Section VIII.

\section{Background}

In this section, we first introduce the notations and the state variables of interest, and then we give the densities/integrals for trajectories and the general multi-object modelling. At last, we present the PMBM density representation and the TPMB approximation. More details on the background can be found in \cite{garcia2019multiple,garcia2020trajectory}.

For a generic space $D$, the set of finite subsets of $D$ is denoted by ${\cal F}(D)$, and the cardinality of a set $A\in {\cal F}(D)$ is $|A|$. We use $\uplus$ to denote the union of sets that are mutually disjoint, $\langle f,g \rangle$ to denote the inner product $\int f(x)g(x) dx$, and the multi-object exponential $f^A$, for some real-valued function $f$, to denote the product $\prod_{x\in A}f(x)$ with $f^{\emptyset} = 1$ by convention. In addition, we use $\delta_x[\cdot]$ and $\delta_x(\cdot)$ to represent the Kronecker delta and the generalized Dirac delta functions centred at $x$, respectively \cite{graham1989concrete,vo2015multitarget}. Here, $x$ in $\delta_x(\cdot)$ belongs to the space of $\nu$ Cartesian products of space $\mathcal{X}$, i.e., $x \in \mathcal{X}^\nu$.

\subsection{State variables}

The single-object state $x_k \in {\cal X}$ at time step $k$, where ${\cal X}$ is a locally compact, Hausdorff and second-countable (LCHS) space \cite[Section 2.2.2]{rfs}, contains information of interest of a single object, including its kinematic state (e.g., position and velocity) and possibly also its extent state, which describes the object shape and size. The set of object states at time step $k$ is an RFS ${\bf x}_k = \{x_k^1,\dots,x_k^{n_k}\} \in {\cal F}({\cal X})$. The measurements collected by the sensor at time step $k$ is ${\bf z}_k = \{z_k^1,\dots,z_k^{m_k}\}$ with single measurement $z_k^j \in \mathbb{R}^{d_z}$ for $j \in \{1,\dots,m_k\}$, and the sequence of measurement sets up to and including time step $k$ is denoted as ${\bf z}_{1:k}$.

The trajectory of an object is a finite sequence of its states at consecutive time steps. In this paper, we follow the notational convention in \cite{garcia2019multiple,garcia2019trajectory,garcia2020trajectory,garcia2020metric} to denote a single trajectory as $X = (t,x^{1:\nu})$ where $t$ is the initial time step of the trajectory, $\nu$ is its length and $x^{1:\nu} = (x^1,\dots,x^\nu)$ is a finite sequence of object states \cite{garcia2019multiple,garcia2019trajectory,garcia2020trajectory,garcia2020metric}. The variable $(t,\nu)$ belongs to the set $I_{(k)} = \{(t,\nu):0\leq t \leq k, 1\leq \nu \leq k-t+1\}$. A single trajectory $X$ up to time step $k$ therefore belongs to the space $T_{(k)} = \uplus_{(t,\nu)\in I_{(k)}}\{t\}\times \mathcal{X}^{\nu}$, which is also LCHS \cite[Appendix A]{garcia2019multiple}. The set of trajectories up to the current time step $k$ is denoted as ${\bf X}_k \in {\cal F}(T_{(k)})$.  

We note that trajectory $X$ is a combination of discrete and continuous states. Such a hybrid state is not uncommon in MOT: a typical example is the interacting multiple model \cite{blom1988interacting}. We also note that the dimension of the state sequence $x^{1:\nu}$ becomes high for long trajectories. This, however, does not necessarily make the computation of multi-object filtering recursions complex at each time step, see, e.g., the efficient MOT implementations in \cite{granstrom2018poisson,garcia2020trajectory}.

\subsection{Densities and integrals}

Given a real-valued function $\pi(\cdot)$ on the single trajectory space $T_{(k)}$, its integral is \cite{garcia2020trajectory} 
\begin{equation}
        \int \pi(X)dX = \sum_{(t,\nu)\in I_{(k)}}\int \pi\left(t,x^{1:\nu}\right)dx^{1:\nu},
\end{equation}
which goes through all possible start times, lengths and object states of the trajectory. Since the single trajectory space $T_{(k)}$ is LCHS, we can apply Mahler's FISST theory set integral \cite[Section 3.3]{rfs} to a space of sets of trajectories ${\cal F}(T_{(k)})$ \cite{xia2019multi}. Specifically, given a real-valued function $f(\cdot)$ on the space ${\cal F}(T_{k})$ of finite sets of trajectories, its set integral is 
\begin{equation}
        \int f({\bf X}_k)\delta {\bf X}_k = \sum_{n=0}^{\infty}\frac{1}{n!}\int f (\{X_1,\dots,X_n\}) d(X_1,\dots,X_n).
\end{equation}
A function $f(\cdot)$ is a multi-trajectory density of a random finite set of trajectories if $f(\cdot)\geq 0$ and its set integral is one. The reference measures and measure theoretic integrals for sets of trajectories are defined in \cite{xia2019multi}.

\subsection{Multi-object modelling}\label{sec_model}

\subsubsection{Multi-object dynamic model}
At time step $k$, new objects appear in the sensor's field of view, following a PPP with birth intensity $\lambda^B_{k}(x_{k})$, independently of any existing objects. For an existing object $x_{k-1}$, it survives with probability $p^S(x_{k-1})$, and if it survives, its state evolves with a Markovian transition density $g_{k}(x_{k}|x_{k-1})$, independently of any other objects. It is also assumed that objects that have disappeared never reappear.

\subsubsection{Multi-object measurement model} \label{sec_clutter}
At time step $k$, the set ${\bf z}_k$ of measurements is the union of a set of object-generated measurements and a set of clutter measurements. The measurements from each object are independent of other objects and of clutter measurements. The set of clutter measurements is a PPP with Poisson intensity $\lambda_k^{C}(z_k) = \gamma_k^C\mu_k^C(z_k)$ where $\gamma_k^C$ is the Poisson rate and $\mu_k^C(z_k)$ is the clutter spatial distribution. 

A single object with state $x_k$ generates an independent set ${\bf w}_k$ of measurements with density $\ell_k({\bf w}_k|x_k)$. Note that in terms of multi-object conjugate priors, the difference between point and extended object tracking is only how the single-object measurement likelihood $\ell_k(\cdot|x_k)$ is defined \cite{pmbmextended2,lmbextended}. For a point object, ${\bf w}_k$ is a Bernoulli RFS with at maximum one measurement, whereas for an extended object, ${\bf w}_k$ is commonly modelled as a PPP with state dependent Poisson rate $\gamma_k(x_k)$ and spatial distribution $\ell_k(\cdot|x_k)$ \cite{ppp}. For a set ${\bf w}_k$ of measurements generated by an extended object with state $x_k$, its Poisson set density is \cite[Section 21.2.4]{rfs}
\begin{equation}\label{eq_ppp_meas}
        \ell_k({\bf w}_k|x_k) = e^{-\gamma_k(x_k)}\prod_{z_k \in {\bf w}_k}\gamma_k(x_k)\ell_k(z_k|x_k),
\end{equation}
and, accordingly, we have the effective probability of detection, i.e., the probability that object $x_k$ generates at least one measurement 
\begin{equation*}
  p(|\mathbf{w}_k|>0|x_k) = 1-\ell_k(\emptyset|x_k) =1-e^{-\gamma_k(x_k)},
\end{equation*}
where $\emptyset$ denotes an empty set.

\subsection{PMBM conjugate prior}

Given the sequence of measurements ${\bf z}_{1:k}$ up to time step $k$ and the multi-object models in Section \ref{sec_model} for point objects where the set of object-generated measurement is a Bernoulli RFS, the density $f_{k|k^\prime}(\cdot)$ of the set of trajectories at time step $k^\prime\in \{k-1,k\}$ is a PMBM \cite{granstrom2018poisson,granstrom2019poisson} with
\begin{align}
  f_{k|k^\prime}({\bf X}_{k}) &= \sum_{{\bf Y}\uplus {\bf V} = {\bf X}_{k}} f^p_{k|k^\prime}({\bf Y})f^{mbm}_{k|k^\prime}({\bf V})\label{eq_pmbm},\\
  f^p_{k|k^\prime}({\bf Y}) &= e^{-\left\langle \lambda_{k|k^\prime},1 \right\rangle} \left[ \lambda_{k|k^\prime}(\cdot) \right]^{{\bf Y}}\label{eq_ppp},\\
  f^{mbm}_{k|k^\prime}({\bf V}) &= \sum_{a\in {\cal A}_{k|k^\prime}}w^a_{k|k^\prime}\sum_{\uplus_{l=1}^{n_{k|k^\prime}}{\bf X}^l = {\bf V}}\prod_{i=1}^{n_{k|k^\prime}}f^{i,a^i}_{k|k^\prime}\left({\bf X}^i\right)\label{eq_mbm},\\
  f^{i,a^i}_{k|k^\prime}\left({\bf X}^i\right) &= \begin{cases}
    1 - r^{i,a^i}_{k|k^\prime} & {\bf X}^i = \emptyset\\
    r^{i,a^i}_{k|k^\prime}f^{i,a^i}_{k|k^\prime}(X) & {\bf X}^i = \{X\}\\
    0 & \text{otherwise},
  \end{cases}\label{eq_ber}
\end{align}
where the prior density at time step $0$ is given by $f_{0|0}(\cdot)$. The multi-object density $f_{k|k^\prime}(\cdot)$ of the form \eqref{eq_pmbm} is shown to be a multi-object conjugate prior \cite{pmbmpoint,granstrom2018poisson}, where it can be observed that the PPP $f^p_{k|k^\prime}(\cdot)$ represents trajectories of objects that are hypothesized to exist but have never been detected\footnote{The set of undetected trajectories arises naturally from the standard multi-object models with Poisson birth, see \cite{pmbmpoint} for further explanations. The PPP of undetected trajectories is independent of the MBM that represents trajectories detected at some point.}, and that the mixture of multi-Bernoulli (MB) components $f^{mbm}_{k|k^\prime}(\cdot)$ represents potential trajectories that have been detected at least once at some point to time step $k^{\prime}$. In \eqref{eq_ppp}, $\lambda_{k|k^\prime}(\cdot)$ denotes the PPP intensity function, whereas in \eqref{eq_mbm} each MB component describes the distribution of detected trajectories conditioned on global association hypothesis $a \in {\cal A}_{k|k^\prime}$, and its weight $w^a_{k|k^\prime}$ represents the probability of the corresponding data association hypothesis \cite{pmbmpoint}. In \eqref{eq_mbm}, there are $n_{k|k^\prime}$ Bernoulli components, indexed by variable $i$, and for each Bernoulli there are $h^i_{k|k^\prime}$ local hypotheses. The density of the $i$-th Bernoulli component with local hypothesis $a^i\in\{1,\dots,h^i_{k|k^\prime}\}$ is given by \eqref{eq_ber}, where $r^{i,a^i}_{k|k^\prime}$ is the probability of existence and $f^{i,a^i}_{k|k^\prime}(\cdot)$ is the single-trajectory density \cite{{granstrom2019poisson}}. A global hypothesis $a = (a^1,\dots,a^{n_{k|k^\prime}})\in {\cal A}_{k|k^\prime}$ selects a local hypothesis for each Bernoulli component, where the selected local hypotheses together form a valid data association and ${\cal A}_{k|k^\prime}$ is the set of global hypotheses that will be described in Section \ref{sec_filtering_recursion}. The weight of global hypothesis $a$ satisfies 
\begin{equation}
  w^a_{k|k^\prime} \propto \prod_{i=1}^{n_{k|k^\prime}}w^{i,a^i}_{k|k^\prime}
\end{equation}
where $w^{i,a^i}_{k|k^\prime}$ is the weight of local hypothesis $a_i$ for the $i$-th Bernoulli component, and normalization is required to ensure that $\sum_{a\in {\cal A}_{k|k^\prime}}w^a_{k|k^\prime} = 1$.

The explicit modelling of undetected objects is beneficial for automotive applications, e.g., in traffic scenes where pedestrians walking on side walks are occluded by parked vehicles along the street. More examples can be found in \cite{granstrom2022tutorial,pmbmextended2}. We also note that MOT algorithms based on sets of trajectories, e.g., the TPMBM filter \cite{granstrom2018poisson} and the TPMB filter \cite{garcia2020trajectory}, do not produce interrupted trajectories as this is an impossible event according to the standard models.

\subsection{Trajectory PMB approximation}

The PMB is a special case of PMBM with a single MB component. We can form PMB approximations using different techniques, see, e.g., \cite{pmbmpoint,garcia2020trajectory,xia2021poisson}. In this work, we follow \cite{garcia2020trajectory} where the best PMB approximation is defined by using Kullback-Leibler divergence (KLD) minimization, on a trajectory space with auxiliary variables \cite{garcia2020trajectory}. The KLD on the space of sets of trajectories with auxiliary variables is an upper bound on the KLD for sets of trajectories without auxiliary variables \cite[Lemma 3]{garcia2020trajectory}.

\subsubsection{PMBM with auxiliary variables}\label{sec_auxiliary}
We extend the single trajectory space with an auxiliary variable $u\in\mathbb{U}_{k|k^\prime}=\{0,1,\dots,n_{k|k^\prime}\}$, such that $(u,X)\in \mathbb{U}_{k|k^\prime}\times T_{(k)}$, and we denote a set of trajectories with auxiliary variables as ${\widetilde{{\bf X}}}_{k}\in {\cal F}(\mathbb{U}_{k|k^\prime}\times T_{(k)})$. Given $f_{k|k^\prime}(\cdot)$ of the form \eqref{eq_pmbm}, the density $\widetilde{f}_{k|k^\prime}(\cdot)$ on the space ${\cal F}(\mathbb{U}_{k|k^\prime}\times T_{(k)})$ can be defined as \cite{garcia2020trajectory}
\begin{equation}\label{eq_pmbm_au}
  \widetilde{f}_{k|k^\prime}\left(\widetilde{{\bf X}}_k\right) = \widetilde{f}^p_{k|k^\prime}\left(\widetilde{{\bf Y}}_k\right)\sum_{a\in{\cal A}_{k|k^\prime}}w^a_{k|k^\prime}\prod_{i=1}^{n_{k|k^\prime}}\widetilde{f}^{i,a^i}_{k|k^\prime}\left(\widetilde{{\bf X}}^i_k\right)
\end{equation}
where, for a given $\widetilde{{\bf X}}_k$, $\widetilde{{\bf Y}}_k = \{(u,X)\in \widetilde{{\bf X}}_k: u =0\}$ and $\widetilde{{\bf X}}^i_k = \{(u,X)\in \widetilde{{\bf X}}_k: u =i\}$, and 
\begin{subequations}
\begin{align}
  \widetilde{f}^p_{k|k^\prime}\left(\widetilde{{\bf Y}}_{k}\right) &= e^{-\left\langle \widetilde{\lambda}_{k|k^\prime},1 \right\rangle} \left[ \lambda_{k|k^\prime}(\cdot) \right]^{\widetilde{{\bf Y}}_{k}},\\
  \widetilde{\lambda}_{k|k^\prime}(u,X) &= \delta_0[u]\widetilde{\lambda}_{k|k^\prime}(X),\\
  \widetilde{f}^{i,a^i}_{k|k^\prime}\left(\widetilde{{\bf X}}_k^i\right) &= \begin{cases}
    1 - r^{i,a^i}_{k|k^\prime} & \widetilde{{\bf X}}_k^i = \emptyset\\
    r^{i,a^i}_{k|k^\prime}f^{i,a^i}_{k|k^\prime}(X)\delta_i[u] & \widetilde{{\bf X}}_k^i = \{(u,X)\}\\
    0 & \text{otherwise}.
  \end{cases}
\end{align}
\end{subequations}
Here, the notation $\widetilde{f}_{k|k^\prime}(\cdot)$ is adopted to clarify that it is the distribution on the augmented space incoporating the auxiliary variables. We also note that the sum over sets in \eqref{eq_pmbm} disappears in \eqref{eq_pmbm_au} due to the use of auxiliary variables as now there is only one possible partition of $\widetilde{{\bf X}}_k$ into $\widetilde{{\bf Y}}_k$, $\widetilde{{\bf X}}^1_k,\dots,\widetilde{{\bf X}}^{n_{k|k^\prime}}_k$ that provides a non-zero density.

\subsubsection{PMB approximation}\label{sec_pmb}
Given a PMBM density $\widetilde{f}_{k|k^\prime}(\cdot)$ of the form \eqref{eq_pmbm_au}, the PMB density that minimizes the KLD from $\widetilde{f}_{k|k^\prime}(\cdot)$ has a closed form and is given by \cite[Proposition 2]{garcia2020trajectory}
\begin{subequations}\label{eq_pmb}
  \begin{align}
    \widetilde{f}^{pmb}_{k|k^\prime}(\widetilde{{\bf X}}_{k}) &= \widetilde{f}^p_{k|k^\prime}(\widetilde{{\bf Y}}_k)\prod_{i=1}^{n_{k|k^\prime}}\widetilde{f}^{i}_{k|k^\prime}\left(\widetilde{{\bf X}}^i_k\right),\label{eq_pmb_predict}\\
    \widetilde{f}^{i}_{k|k^\prime}(\widetilde{{\bf X}}) &= \begin{cases}
      1 - r^{i}_{k|k^\prime} & \widetilde{{\bf X}} = \emptyset\\
      r^{i}_{k|k^\prime}f^{i}_{k|k^\prime}(X)\delta_i[u] & \widetilde{{\bf X}} = \{(u,X)\}\\
      0 & \text{otherwise},
    \end{cases}\\
    r^i_{k|k^\prime} &= \sum_{a^i=1}^{h^i_{k|k^\prime}}\overline{w}_{k|k^\prime}^{i,a^i}r^{i,a^i}_{k|k^\prime},\label{eq_pmb_r}\\
    f^i_{k|k^\prime}(X) &= \frac{\sum_{a^i=1}^{h^i_{k|k^\prime}}\overline{w}_{k|k^\prime}^{i,a^i}r^{i,a^i}_{k|k^\prime}f_{k|k^\prime}^{i,a^i}(X)}{\sum_{a^i=1}^{h^i_{k|k^\prime}}\overline{w}_{k|k^\prime}^{i,a^i}r^{i,a^i}_{k|k^\prime}},\label{eq_pmb_f}\\
    \overline{w}_{k|k^\prime}^{i,a^i} &= \sum_{b\in {\cal A}_{k|k^\prime}:b^i=a^i}w^b_{k|k^\prime}\label{eq_marginal_assoc}.
  \end{align}
\end{subequations}
If we integrate out the auxiliary variables in \eqref{eq_pmb}, we obtain the PMB density without auxiliary variables \cite{garcia2020trajectory}
\begin{equation}\label{eq_pmb_original}
  {f}^{pmb}_{k|k^\prime}({{\bf X}}_{k}) = \sum_{\uplus_{l=1}^{n_{k|k^\prime}}{\bf X}^l\uplus {\bf Y} = {\bf X}_k} {f}^p_{k|k^\prime}({{\bf Y}})\prod_{i=1}^{n_{k|k^\prime}}{f}^{i}_{k|k^\prime}\left({{\bf X}}^i\right)
\end{equation}
where the $i$-th Bernoulli component $f^i_{k|k^\prime}({\bf X}^i)$ is parameterized by $r^i_{k|k^\prime}$ in \eqref{eq_pmb_r} and $f^i_{k|k^\prime}(X)$ in \eqref{eq_pmb_f}.

\section{Problem Formulation and Filtering Recursions}

In this section, we present the problem formulations and the PMBM filtering recursions based on the multi-object models introduced in Section \ref{sec_model} with a generalized measurement model.

\subsection{Problem formulation}

We consider the following two MOT problem formulations:
\begin{enumerate}
  \item The set ${\bf X}_k$ of alive trajectories that exist at the current time step, i.e., $t+\nu-1=k$ for each $(t,x^{1:\nu}) \in {\bf X}_k$.
  \item The set ${\bf X}_k$ of all trajectories that have existed up to the current time step $k$, i.e., $t+\nu-1\leq k$ for each $(t,x^{1:\nu})\in {\bf X}_k$.
\end{enumerate}
It should be noted that the set of trajectories included in both 1) and 2) does not depend on the sensor model, but on the object dynamic model in Section \ref{sec_model}. Therefore, trajectories of objects that move outside the sensor's field-of-view, and appear in the field-of-view afterwards will always belong to 2), and will belong to 1) if the dynamic model indicates they are alive.

For both problem formulations, the objective is to compute the posterior density of the set ${\bf X}_k$ of trajectories at time step $k$ given the sequence ${\bf z}_{1:k}$ of measurements up to time step $k$ using Bayesian recursive filtering. 

\subsection{Bayesian models for sets of trajectories}\label{sec_multitra_model}
We proceed to give the Bayesian multi-trajectory dynamic and measurement models for the two types of problem formulations (see also \cite{garcia2019multiple,garcia2020trajectory}), required for deriving the Bayesian filtering recursions. The two multi-trajectory dynamic models extend the multi-object dynamic model in Section \ref{sec_model}, where each object evolves according to a Markov process, to trajectories.

\subsubsection{Dynamic model for sets of alive trajectories}\label{sec_dynamic_alive}
Given the set ${\bf X}_{k-1}$ of alive trajectories at time step $k-1$, each $X = (t,x^{1:\nu}) \in {\bf X}_{k-1}$ survives with probability $p^S(X) = p^S(x^{\nu})$ (i.e., the survival probability depends only on the final state in the trajectory), and if it survives, its state $(t,x^{1:\nu})$ evolves with a transition density \cite[Sec. IV-A-1]{garcia2020trajectory}
\begin{align}
  g_{k}\left(\bar{t},y^{1:\bar{\nu}}|X\right) &= \delta_t[\bar{t}]\delta_{\nu+1}[\bar{\nu}]\delta_{x^{1:\nu}}\left(y^{1:\bar{\nu}-1}\right)g_{k}\left(y^{\bar{\nu}}|x^\nu\right),
\end{align}
where the state $y^{\bar{\nu}}$ only depends on $x^\nu$ due to the Markov property.

The set ${\bf X}_{k}$ is the union of the surviving trajectories and the PPP newborn trajectories with birth intensity 
\begin{equation}
  \label{eq_birth}
  \lambda^B_{k}\left(t,x^{1:\nu}\right) = \delta_{k}[t]\delta_1[\nu]\lambda^B_{k}\left(x^\nu\right).
\end{equation}

\subsubsection{Dynamic model for sets of all trajectories}\label{sec_dynamic_all}
Given the set ${\bf X}_{k-1}$ of all trajectories at time step $k-1$, each $X = (t,x^{1:\nu}) \in {\bf X}_{k-1}$ survives with probability $p^S(X) = 1$ and evolves with a transition density \cite[Sec. IV-A-2]{garcia2020trajectory}
\begin{align}
  &g_{k}\left(\bar{t},y^{1:\bar{\nu}}|X\right) = \delta_t[\bar{t}]\nonumber\\
  &\times \begin{cases}
    \delta_{\nu}[\bar{\nu}]\delta_{x^{1:\nu}}\left(y^{1:\bar{\nu}}\right) & \bar{\omega} < k-1\\
    \delta_\nu[\bar{\nu}]\delta_{x^{1:\nu}}\left(y^{1:\bar{\nu}}\right)\left(1-p^S\left(x^\nu\right)\right) & \bar{\omega} = k-1\\
    \delta_{\nu+1}[\bar{\nu}]\delta_{x^{1:\nu}}\left(y^{1:\bar{\nu}-1}\right)p^S\left(x^\nu\right) g_{k}\left(y^{\bar{\nu}}|x^\nu\right) & \bar{\omega} = k
  \end{cases}
\end{align}
where $\bar{\omega} = \bar{t} + \bar{\nu} -1$ and the state $y^{\bar{\nu}}$ only depends on $x^\nu$ due to the Markov property. The birth model is also a PPP with intensity \eqref{eq_birth}. Note that $p^S(X)$ refers to the probability that a trajectory remains in the considered set of trajectories for both problem formulations. 

\subsubsection{Measurement model for sets of trajectories}
The measurement model is the same for sets of alive trajectories and the sets of all trajectories. Each trajectory $(t,x^{1:\nu})\in {\bf X}_k$ generates a set ${\bf w}_k$ of measurements with density $\ell_k({\bf w}_k|t,x^{1:\nu}) = \ell_k({\bf w}_k|x^\nu)$. The clutter model is a PPP as described in Section \ref{sec_clutter}.

\subsection{Filtering recursions}\label{sec_filtering_recursion}

We present the prediction and update for the multi-trajectory dynamic and measurement models in Section \ref{sec_multitra_model}.

\begin{proposition}\label{pmbm_prediction}
Given the PMBM filtering density on the set of trajectories at time step $k-1$ of the form \eqref{eq_pmbm}, the predicted density at time step $k$ is a PMBM of the form \eqref{eq_pmbm}, with
\begin{subequations}
  \begin{align}
    \lambda_{k|k-1}(X) &= \lambda^B_k(X) + \left\langle \lambda_{k-1|k-1},g_k(X|\cdot)p^S(\cdot) \right\rangle,\\
    n_{k|k-1} &= n_{k-1|k-1},\\
    h^i_{k|k-1} &= h^i_{k-1|k-1},\\
    w^{i,a^i}_{k|k-1} &= w^{i,a^i}_{k-1|k-1},\\
    r^{i,a^i}_{k|k-1} &= r^{i,a^i}_{k-1|k-1}\left\langle f^{i,a^i}_{k-1|k-1},p^S \right\rangle,\\
    f^{i,a^i}_{k|k-1}(X) &= \frac{\left\langle f^{i,a^i}_{k-1|k-1},g_k(X|\cdot)p^S(\cdot) \right\rangle}{\left\langle f^{i,a^i}_{k-1|k-1},p^S \right\rangle}
  \end{align}
\end{subequations}
where $g_k(\cdot|\cdot)$ and $p^S(\cdot)$ are chosen depending on the problem formulation.
\end{proposition}

The TPMBM prediction given in Proposition \ref{pmbm_prediction} is the same as the one in \cite{granstrom2018poisson} as it is not affected by the choice of the measurement model.

Before presenting the update step, it is important to define the set of feasible global hypotheses for a generalized measurement model. We refer to measurement $z_k^j$ using the pair $(k,j)$ and the set of all such measurement pairs up to (and including) time step $k$ is denoted by ${\cal M}_k$. Then, a local hypothesis $a^i$ for the $i$-th Bernoulli component has a set of measurement pairs denoted as ${\cal M}_k^{i,a^i} \subseteq {\cal M}_k$, and the set ${\cal A}_{k|k^\prime}$ of all global hypotheses satisfies
\begin{multline}\label{eq_globalhypo}
  {\cal A}_{k|k^\prime} = \left\{ \left(a^1,\dots,a^{n_{k|k^\prime}}\right):a^i \in \left\{1,\dots,h^i_{k|k^\prime}\right\}~\forall~i,\right.\\ \left. \biguplus_{i=1}^{n_{k|k^\prime}}{\cal M}_{k^\prime}^{i,a^i} = {\cal M}_{k^\prime} \right\}
\end{multline}
where the recursive constructions of ${\cal M}_{k^\prime}^{i,a^i}$ and ${\cal M}_{k^\prime}$, with ${\cal M}_{0} = \emptyset$, will be given in Proposition \ref{prop_update}. 

We note that every global hypothesis needs to explain the association of every measurement that has been received so far. The same measurement cannot be associated to more than one local hypothesis, but more than one measurement may be associated to the same local hypothesis at the same time step. Each global hypothesis therefore corresponds to a unique partition of ${\cal M}_{k^\prime}$, and the number of global hypotheses is then given by the Bell number of $|{\cal M}_{k^\prime}|$ \cite{pmbmextended2}. At each time step, each measurement generates a new Bernoulli component, corresponding to an undetected object detected for the first time or clutter.

\begin{proposition}\label{prop_update}
Given the PMBM predicted density on the set of trajectories at time step $k$ of the form \eqref{eq_pmbm}, and measurements ${\bf z}_k = \{z_k^1,\dots,z_k^{m_k}\}$, the updated density is a PMBM of the form \eqref{eq_pmbm}, with
\begin{align}
  n_{k|k} &= n_{k|k-1} + m_k\label{eq_n},\\
  {\cal M}_k &= {\cal M}_{k-1} \cup \left\{ (k,j) | j\in \{1,\dots,m_k\} \right\},\\
  \lambda_{k|k}(X) &= \ell_k(\emptyset|X)\lambda_{k|k-1}(X).
\end{align}
For Bernoulli components continuing from previous time steps $i\in \{1,\dots,n_{k|k-1}\}$, a new local hypothesis is included for each previous local hypothesis and either a misdetection or an update with a non-empty subset of ${\bf z}_k$. The updated number of local hypotheses is $h^i_{k|k} = 2^{m_k}h^i_{k|k-1}$. For missed detection hypotheses, $i\in \{1,\dots,n_{k|k-1}\}$, $a^i\in\{1,\dots,h^i_{k|k-1}\}$, we have
\begin{subequations}
  \begin{align}
    {\cal M}_k^{i,a^i} &= {\cal M}_{k-1}^{i,a^i},\\
    \ell_{k|k}^{i,a^i,0} &= \left\langle f_{k|k-1}^{i,a^i},\ell_k(\emptyset|\cdot) \right\rangle,\\
    w_{k|k}^{i,a^i} &= w^{i,a^i}_{k|k-1}\left(1 - r^{i,a^i}_{k|k-1} + r^{i,a^i}_{k|k-1}\ell_{k|k}^{i,a^i,0}\right),\\
    r^{i,a^i}_{k|k} &= \frac{r^{i,a^i}_{k|k-1}\ell_{k|k}^{i,a^i,0}}{1 - r^{i,a^i}_{k|k-1} + r^{i,a^i}_{k|k-1}\ell_{k|k}^{i,a^i,0}},\\
    f^{i,a^i}_{k|k}(X) &= \frac{\ell_k(\emptyset|X)f^{i,a^i}_{k|k-1}(X)}{\ell_{k|k}^{i,a^i,0}}.
  \end{align}
\end{subequations}
Let ${\bf w}_k^1,\dots,{\bf w}_k^{2^{m_k}-1}$ be the non-empty subsets of ${\bf z}_k$. For the $i$-th Bernoulli component, $i\in\{1,\dots,n_{k|k-1}\}$, with a local hypothesis $\tilde{a}^i\in \{1,\dots,h^i_{k|k-1}\}$ in the predicted density, the new local hypothesis generated by a set ${\bf w}^j$ of measurements has $a^i = \tilde{a}^i + h^i_{k|k-1}j$ with $j\in\{1,\dots,2^{m_k}-1\}$, and 
\begin{subequations}
  \begin{align}
    {\cal M}_k^{i,a^i} &= {\cal M}_{k-1}^{i,\tilde{a}^i} \cup \left\{ (k,p):z_k^p \in {\bf w}_k^j \right\},\\
    \ell_{k|k}^{i,a^i,j} &= \left\langle f_{k|k-1}^{i,\tilde{a}^i},\ell_k\left({\bf w}^j_k|\cdot\right) \right\rangle,\\
    w_{k|k}^{i,a^i} &= w_{k|k-1}^{i,\tilde{a}^i}r_{k|k-1}^{i,\tilde{a}^i}\ell_{k|k}^{i,a^i,j},\\
    r_{k|k}^{i,a^i} &= 1,\\
    f_{k|k}^{i,a^i}(X) &= \frac{\ell_k\left({\bf w}_k^j|X\right)f_{k|k-1}^{i,\tilde{a}^i}(X)}{\ell_{k|k}^{i,a^i,j}}.
  \end{align}
\end{subequations}
Each new Bernoulli component has a different number of local hypotheses, each of which is created by a subset of ${\bf z}_k$.  Let the set ${\cal S}_i$ of subsets of measurements associated to the $i$-th new Bernoulli component ($i\in \{1,\dots,m_k\}$) be recursively built as
\begin{equation}\label{eq_S}
  {\cal S}_i = \left\{ \left\{ {z}_k^i \right\} \right\} \cup \left( \bigcup_{{\bf w}\in \cup_{j=1}^{i-1}{\cal S}_i} \left\{ \left\{ { z}_k^i \right\} \cup {\bf w}  \right\} \right)
\end{equation}
with ${\cal S}_1 = \{ \{ {z}_k^1 \} \}$, where $\{ \{ {z}_k^i \} \}$ denotes a set of subsets, containing the single element set $\{z^i_k\}$. According to \eqref{eq_S}, the $i$-th new Bernoulli component has $2^{i-1}$ local hypotheses that are created by non-empty subsets of measurements.
Further, let ${\bf w}_k^{i,\iota}$ denote the $\iota$-th subset of measurements of the $i$-th Bernoulli component ($i = n_{k|k-1}+j,j\in\{1,\dots,m_k\},\iota\in \{1,\dots,2^{j-1}\}$). Then, for the $i$-th Bernoulli component, there are $h^i_{k|k} = 2^{j-1}+1$ local hypotheses, one corresponding to a non-existent Bernoulli component ($a^i = 1$)
\begin{equation}\label{eq_newBernoulli1}
  {\cal M}_k^{i,1} = \emptyset,w_{k|k}^{i,1} = 1,r_{k|k}^{i,1} = 0,
\end{equation}
and the others ($a^i = \iota+1,\iota\in \{1,\dots,2^{j-1}\}$) have
\begin{subequations}\label{eq_newBernoulli2}
  \begin{align}
    {\cal M}_k^{i,a^i} &= \left\{ (k,p):z_k^p\in {\bf w}_k^{i,\iota} \right\},\\
    \ell_{k|k}^{i,\iota} &= \left\langle \lambda_{k|k-1},\ell_k\left({\bf w}_k^{i,\iota}|\cdot\right) \right\rangle,\\
    w_{k|k}^{i,a^i} &= \delta_1\left[\left|{\bf w}_k^{i,\iota}\right|\right]\left[\lambda_k^C\right]^{{\bf w}_k^{i,\iota}} + \ell_{k|k}^{i,\iota},\\
    r_{k|k}^{i,a^i} &= \frac{\ell_{k|k}^{i,\iota}}{w_{k|k}^{i,a^i}},\\
    f_{k|k}^{i,a^i}(X) &= \frac{\ell_k\left({\bf w}_k^{i,\iota}|X\right)\lambda_{k|k-1}(X)}{\ell_{k|k}^{i,\iota}}.
  \end{align}
\end{subequations}
\end{proposition}

\begin{figure}[!t]
  \centering
  \includegraphics[width=\columnwidth]{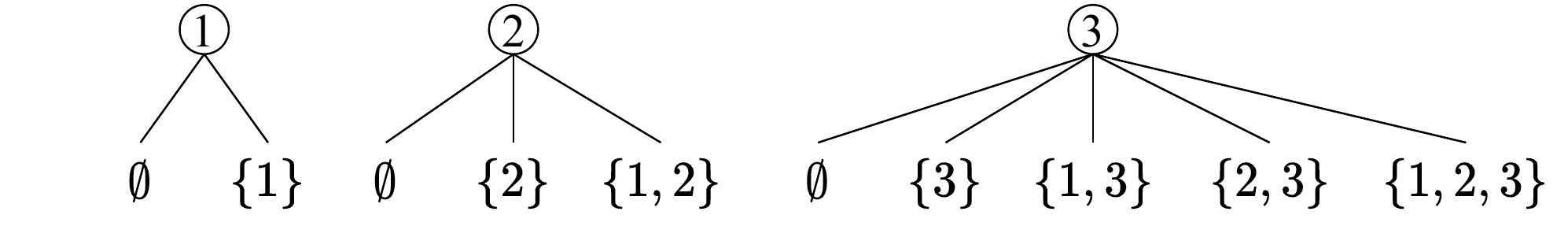}
  \caption{An illustration of the local hypothesis structure in Proposition \ref{prop_update} for the case $n_{k|k-1} = 0$ and $m_k=3$ where the time step $k$ is omitted. In this case, three Bernoulli components are created, one for each measurement. The local hypotheses under these components are: 1) $\emptyset$ and $\{(k,1)\}$, 2) $\emptyset$, $\{(k,2)\}$ and $\{(k,1),(k,2)\}$, 3) $\emptyset$, $\{(k,3)\}$, $\{(k,1),(k,3)\}$, $\{(k,2),(k,3)\}$ and $\{(k,1),(k,2),(k,3)\}$. }
  \label{fig_pmb_example}
\end{figure}

Proposition \ref{prop_update} is an extension of the PMBM update for sets of objects  \cite[Theorem 1]{garcia2021poisson} to sets of trajectories, and a short proof of Proposition \ref{prop_update} is given in Appendix \ref{appendix_a}.

Compared to the local hypothesis representation in \cite[Theorem 1]{garcia2021poisson}, in Proposition \ref{prop_update} each measurement, instead of each non-empty subset of measurements, creates a new Bernoulli component, and the $i$-th new potential object cannot generate more than $i$ measurements, see Fig. \ref{fig_pmb_example} for an example. These two different local hypothesis representations lead to the same PMBM posterior, but the one in Proposition \ref{prop_update} facilitates the development of efficient PMB approximation methods \cite{xia2021poisson}. 

The established PMBM conjugacy on posterior of sets of trajectories for a generalized object measurement model paves the way for developing efficient inference methods on sets of trajectories using more general object measurement models for both online and offline applications, using, e.g., BP \cite{meyer2018message} and Gibbs sampling \cite{fatemi2017poisson}. A typical example is the extended object trajectory PMBM filter presented in our preliminary work \cite{xia2019extended}. In the next section, we will elaborate on how to derive the factor graph formulation of the update step by making use of Proposition \ref{prop_update}, with the objective to later develop BP algorithm for online MOT. 

\section{Factor Graph Formulation}

The TPMB filters are computationally lighter alternatives to the TPMBM filters. The prediction of the TPMB filter is a special case of the TPMBM prediction with only one mixture component. The update of the TPMB filter is obtained by first performing a Bayesian update, which yields a PMBM distribution, followed by a PMB approximation. The main challenge of the PMB approximation is the efficient calculation of the marginal density of set of trajectories, where the uncertainty on global hypotheses has been marginalized out. To address this problem, the existing works \cite{garcia2021poisson,xia2021poisson} first prune global hypotheses with negligible weights using either C\&A \cite{xia2019extended,pmbmextended2} or sampling-based methods \cite{soextended}, and then approximately compute the marginal association probabilities \eqref{eq_marginal_assoc} by enumerating the truncated global hypotheses. Also, we note that, even if one manages to compute \eqref{eq_marginal_assoc}, it may still be computationally challenging to evaluate \eqref{eq_pmb_f} in closed form, see, e.g., \cite{gammareduction,thormann2021fusion}.

The present work is motivated by an emergent approximation method for multiple EOT with Poisson spatial model that is both accurate and scalable \cite{florian2021scalable}. The approximation is based on BP, which jointly computes the marginal density of object states as well as the marginal association probability of each measurement, without explicitly enumerating the associations between measurement clusters and objects. To apply the techniques in \cite{florian2020scalable,florian2021scalable} to PMB approximation, we need to represent the joint posterior of set of trajectories and association variables using a factor graph. 

In this section, we give the factor graph representation of the joint posterior of sets of trajectories and association variables, required for developing efficient TPMB filters with Poisson spatial model using BP. 

\subsection{Factor graph formulation}\label{sec_factor_graph_formulation}
 
\begin{figure}[!t]
  \centering
  \includegraphics[width=\columnwidth]{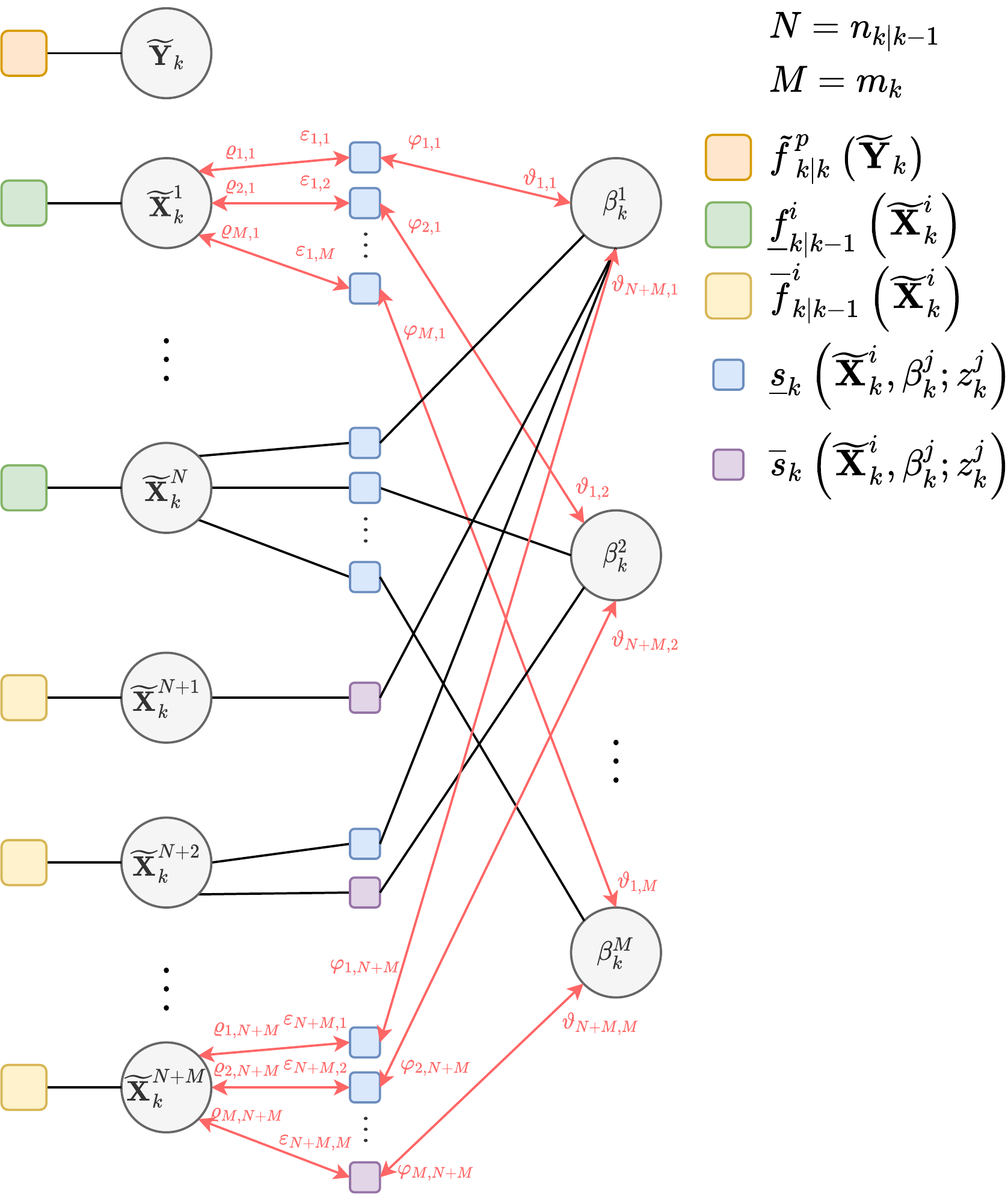}
  \caption{Factor graph for the factorization in \eqref{eq_factor_graph2} where variable nodes are represented using square and factor nodes are represented using circle. The messages passing in loopy BP via $\widetilde{{\bf X}}_k^1$ and $\widetilde{{\bf X}}_k^{n_{k|k-1}+m_k}$ are also illustrated (in red lines) where their arguments and iteration indices are omitted for brevity.}
  \label{fig_factor_graph2}
\end{figure}

Following \cite{florian2021scalable}, we use the measurement-oriented association vector $\beta_k = [\beta^1_k,\dots,\beta_k^{m_k}]^T$ where $\beta_k^j\in\{0,1,\dots,n_{k|k}\}$ for $j\in\{1,\dots,m_k\}$, to describe the object-to-measurement associations. Specifically, $\beta_k^j=i>0$ if and only if the $j$-th measurement $z_k^j$ is associated to the $i$-th potential object (described by the $i$-th Bernoulli component), whereas $\beta_k^j=0$ if the $j$-th measurement is a false alarm and thus not associated to a potential object. According to the local hypothesis representation for new Bernoulli components in Proposition \ref{prop_update}, the $j$-th measurement $z_k^j$ cannot be associated to new potential objects with index $i\in\{ n_{k-1|k}+1,\dots, n_{k-1|k}+j-1\}$, i.e., $\beta_k^j\in\{0,1,\dots,n_{k-1|k},n_{k-1|k}+j,\dots,n_{k|k}\}$.

\begin{proposition}\label{proposition_factor_graph}
  Given a predicted TPMB density at time step $k$ of the form \eqref{eq_pmb_predict}, measurement set ${\bf z}_k$, and the PPP measurement likelihood
  \begin{subequations}\label{eq_measliktra}
    \begin{align}
      \ell_k\left({\bf w}_k|X\right) &= e^{-\gamma_k(X)}\prod_{z_k \in {\bf w}_k}\gamma_k(X)\ell_k(z_k|X),\\
      \gamma_k\left(t,x^{1:\nu}\right) &= \gamma_k\left(x^\nu\right)\delta_k[t+\nu-1],\\
      \ell_k\left(z_k|t,x^{1:\nu}\right) &= \ell_k\left(z_k|x^\nu\right)\delta_k[t+\nu-1],
    \end{align}
  \end{subequations}
  the joint posterior of the set of trajectories and measurement-oriented association vector is 
  \begin{align}\label{eq_factor_graph2}
    &\widetilde{f}_{k|k}\left(\widetilde{{\bf X}}_k,\beta_k\right)\nonumber\\
    &\propto \widetilde{f}^p_{k|k}\left(\widetilde{{\bf Y}}_k\right)\prod_{i=1}^{n_{k|k-1}}\Biggl[\underline{f}^{i}_{k|k-1}\left(\widetilde{{\bf X}}^i_k\right)\prod_{j=1}^{m_k}\underline{s}_k\left(\widetilde{{\bf X}}_k^i,\beta_k^{j};z_k^j\right) \Biggr]\nonumber\\
    &~~\times\prod_{i=n_{k|k-1}+1}^{n_{k|k}}\Biggl[  \overline{f}^i_{k|k-1}\left(\widetilde{{\bf X}}^i_k\right)\overline{s}_k\left(\widetilde{{\bf X}}_k^i,\beta_k^{i-n_{k|k-1}};z_k^{i-n_{k|k-1}}\right)\nonumber\\
    &~~\times\prod_{j=1}^{i-n_{k|k-1}-1}\underline{s}_k\left(\widetilde{{\bf X}}_k^i,\beta_k^{j};z_k^j\right)\Biggr]
  \end{align}
  where $\widetilde{\bf Y}_k$ and $\widetilde{\bf X}_k^i$ with $i\in\{1,\dots,n_{k|k}\}$ are defined as in \eqref{eq_pmbm_au}, and
  \begin{align}
    &\underline{f}^{i}_{k|k-1}\left(\widetilde{{\bf X}}\right)\nonumber\\ &= \begin{cases}
      r^i_{k|k-1}f^i_{k|k-1}(X)e^{-\gamma_k(X)}\delta_i[u] & \widetilde{{\bf X}} = \{(u,X)\}\\
      1 - r^i_{k|k-1} & \widetilde{{\bf X}} = \emptyset\\
      0 & \text{otherwise},
    \end{cases}\label{eq_existingprior}\\
    &\overline{f}^i_{k|k-1}\left(\widetilde{{\bf X}}\right)\nonumber\\ &= \begin{cases}
      \lambda_{k|k-1}(X)e^{-\gamma_k(X)}\delta_i[u] & \widetilde{{\bf X}} = \{(u,X)\}\\
      1 & \widetilde{{\bf X}} = \emptyset\\
      0 & \text{otherwise},
    \end{cases}\label{eq_newbornprior}\\
    &\underline{s}_k\left(\widetilde{{\bf X}}_k^i,\beta_k^j;z_k^j\right)\nonumber\\
    &= \begin{cases}
      \frac{\ell_k\left(z_k^j|X\right)\gamma_k(X)}{\lambda_k^C\left(z_k^j\right)}\delta_i[u] & \widetilde{{\bf X}}_k^i = \{(u,X)\},\beta_k^{j}=i\\
      1 & \beta_k^{j} \neq i\\
      0 & \text{otherwise},
    \end{cases}\label{eq_s_underline}\\
    &\overline{s}_k\left(\widetilde{{\bf X}}_k^i,\beta_k^j;z_k^j\right)\nonumber\\ 
    &= \begin{cases}
      \frac{\ell_k\left(z_k^j|X\right)\gamma_k(X)}{\lambda_k^C\left(z_k^j\right)}\delta_i[u] & \widetilde{{\bf X}}_k^i = \{(u,X)\},\beta_k^{j} = i\\
      1 & \widetilde{{\bf X}}_k^i = \emptyset,\beta_k^{j} \neq i\\
      0 & \text{otherwise}.
    \end{cases}\label{eq_s_overline}
  \end{align}
\end{proposition}

The expression of the joint posterior \eqref{eq_factor_graph2} consists of three factors, including $\widetilde{f}^p_{k|k}(\cdot)$ and two products. The factor $\widetilde{f}^p_{k|k}(\cdot)$ describes the posterior density of the set of undetected trajectories. The first product over $i\in\{1,\dots,n_{k|k-1}\}$ describes the joint posterior of the set of existing trajectories and the measurement-oriented association vector. The second product over $i\in\{n_{k|k-1}+1,\dots,n_{k|k}\}$ describes the joint posterior of the set of newly detected trajectories and the measurement-oriented association vector. 

Each factor in the two products in \eqref{eq_factor_graph2} can be further factorized as a product over several smaller factors. In particular, factors \eqref{eq_existingprior} and \eqref{eq_newbornprior} describe the set $\widetilde{\bf X}^i_k$ of trajectories prior without relating to any measurements, respectively for existing and newly detected objects. Factors \eqref{eq_s_underline} and \eqref{eq_s_overline} reflect the likelihood of associating a particular measurement with its corresponding Bernoulli component, respectively for existing and newly detected objects. We note that an object may have several associated measurements; the factorized likelihood in \eqref{eq_ppp_meas} results in the factorized form of \eqref{eq_factor_graph2}.

Proposition \ref{proposition_factor_graph} is proved in Appendix \ref{appendix_factor_graph}, and the factor graph corresponding to \eqref{eq_factor_graph2} is illustrated in Fig. \ref{fig_factor_graph2}. By leveraging Proposition \ref{prop_update} and RFSs, we can have a concise derivation of the factor graph formulation of the joint posterior, which is precisely built upon the multi-object models given in Section \ref{sec_model} and the Poisson spatial measurement model \eqref{eq_measliktra} without any further assumptions, as compared to the long and involved derivation given in \cite[Section 1]{meyer2021scalable}.

\subsection{Relation to the factor graph in \cite{florian2021scalable}}
\label{sec_factor_graph_florian}
The factor graph formulation in \cite{florian2021scalable} is for the joint posterior of set of object states and measurement-oriented association variables, derived using random vectors. As a comparison, in this work, we consider inference on sets of trajectories and the factor graph formulation is derived using RFSs. Also note that the factor graph in \cite{florian2021scalable} also includes the time evolution of multi-object dynamics but with the assumption that the object survival probability $p^S$ is state-independent, whereas the factor graph in Fig. \ref{fig_factor_graph2} only focuses on the update step. It is also possible to derive a factor graph formulation that considers the joint posterior of set of trajectories and association variables until the current time step, but this is not required for the development of TPMB filters using BP.

Another important difference is that in the factor graph shown in Fig. \ref{fig_factor_graph2} we have one additional variable node and factor node for the set of undetected trajectories, which is a direct result of the explicit modelling of undetected trajectories in the PMBM conjugate prior. This is not the case in \cite{florian2021scalable}, where it is assumed that newborn objects are detected with probability one such that the posterior of the set of undetected objects is zero, and that the prior distribution for newborn objects (cf. \eqref{eq_newbornprior}) needs to be normalized by $1-\ell_k(\emptyset|\cdot)$ to exclude the case that a newborn object may generate zero measurement. The difference on the predicted number of undetected objects at time step $k$ is further elaborated in Appendix \ref{appendix_c}. 

Note that both factor graph formulations are not unique. The one in \cite{florian2021scalable} depends on the mapping between measurements and new potential objects, and a different mapping may result in a different factorization of the posterior. The factorization in \eqref{eq_factor_graph2} depends on the local hypothesis representation for new Bernoulli components since there is a one-to-one mapping between global hypothesis and measurement-oriented association vector. There is a family of hypothesis representations of the posterior that only differ in the representation of new Bernoulli components \cite[Theorem 1]{xia2021poisson}. Different hypothesis representations for the new Bernoulli components yield different joint posteriors on the space of auxiliary variables, and therefore the corresponding factorizations are also different. 

\section{Loopy Belief Propagation for EOT}\label{sec_lbp}

The objective is to compute the marginal density of each Bernoulli set $\widetilde{{\bf X}}^i_k$ of trajectory with $i\in\{1,\dots,n_{k|k}\}$. Since the factor graph in Fig. \ref{fig_factor_graph2} has cycles, we apply loopy BP for inference, where at each time step, messages are computed and processed in parallel \cite{florian2021scalable}. Loopy BP has been applied to solve multiple point object tracking problems \cite{meyer2018message}. Since the factor graph formulations are different for point and extended objects, the message passing equations are also different. In this section, we present the generic equations for message passing and belief calculation for the factor graph in Fig. \ref{fig_factor_graph2}. These equations resemble the message passing equations in \cite{florian2021scalable}, but they are derived using RFSs. For notational brevity, we omit the time index for all the messages.

\subsection{Iterative message passing}
The notations for the different messages are given as follows. At iteration $p\in\{1,\dots,P\}$, we denote the message from variable node $\widetilde{{\bf X}}_k^i$, $i\in\{1,\dots,n_{k|k}\}$, to factor node $\overline{s}_k(\widetilde{{\bf X}}_k^i,\beta_k^j;z_k^j)$ or $\underline{s}_k(\widetilde{{\bf X}}_k^i,\beta_k^j;z_k^j)$, $j\in\{1,\dots,m_k\}$, by $\varepsilon_{i,j}^{(p)}(\cdot)$ and the message in the reverse direction by $\varrho_{j,i}^{(p)}(\cdot)$. In addition, we also denote the message from factor node $\overline{s}_k(\widetilde{{\bf X}}_k^i,\beta_k^j;z_k^j)$ or $\underline{s}_k(\widetilde{{\bf X}}_k^i,\beta_k^j;z_k^j)$ to variable node $\beta_k^j$ by $\vartheta_{i,j}^{(p)}(\cdot)$ and the message in the reverse direction by $\varphi_{j,i}^{(p)}(\cdot)$. 

\subsubsection{Initialization}
For $p=1$, we set $\varepsilon_{i,j}^{(1)}(\cdot) = \underline{f}^i_{k|k-1}(\cdot)$ for $i\in\{1,\dots,n_{k|k-1}\}$ and $\varepsilon_{i,j}^{(1)}(\cdot) = \overline{f}^i_{k|k-1}(\cdot)$ for $i\in\{n_{k|k-1}+1,\dots,n_{k|k}\}$, where $n_{k|k}$ is given by \eqref{eq_n}. Further, we introduce $\varepsilon^{(p)}_{i,j} = \int \varepsilon^{(p)}_{i,j}\left(\{X\}\right)d X + \varepsilon^{(p)}_{i,j}(\emptyset)$, where for $p=1$, $i\in\{1,\dots,n_{k|k-1}\}$, we have
\begin{align}
  \varepsilon^{(1)}_{i,j} &= \int \underline{f}^{i}_{k|k-1}\left(\widetilde{{\bf X}}\right) \delta \widetilde{{\bf X}}\nonumber\\
  &= \sum_{u\in\mathbb{U}_{k|k}}\int \underline{f}^{i}_{k|k-1}(\{(u,X)\}) d X  + \underline{f}^{i}_{k|k-1}(\emptyset)\nonumber\\
  &=r^i_{k|k-1}\int f^i_{k|k-1}(X)e^{-\gamma_k(X)} dX + 1-r^i_{k|k-1},
\end{align}
and for $i\in\{n_{k|k-1}+1,\dots,n_{k|k}\}$,
\begin{align}
  \varepsilon^{(1)}_{i,j} &= \int \overline{f}^i_{k|k-1}\left(\widetilde{{\bf X}}\right) \delta \widetilde{{\bf X}} \nonumber\\
  &=\sum_{u\in\mathbb{U}_{k|k}}\int \overline{f}_{k|k-1}^i(\{(u,X)\}) d X +  \overline{f}_{k|k-1}(\emptyset)\nonumber\\
  &=\int \lambda_{k|k-1}(X)e^{-\gamma_k(X)} d X + 1.
\end{align}

By comparing the initialization step with the update of local hypotheses in Proposition \ref{prop_update}, it is not difficult to observe that the initialization of message $\varepsilon_{i,j}^{(1)}(\cdot)$, $i\in\{1,\dots,n_{k|k-1}\}$, corresponds to the unnormalized misdetection hypothesis density for the $i$-th Bernoulli component, and that the initialization of message $\varepsilon_{i,j}^{(1)}(\cdot)$, $i\in\{n_{k|k-1},\dots,n_{k|k}\}$, corresponds to the unnormalized density of the second local hypothesis of the $i$-th Bernoulli component but with $\ell_k(\emptyset|\cdot)$. 

\subsubsection{Measurement evaluation}
We first describe the messages $\vartheta_{i,j}^{(p)}(\cdot)$, with $i\in\{1,\dots,n_{k|k-1}\}$, $j\in\{1,\dots,m_k\}$ and $i\in\{n_{k|k-1}+1,\dots,n_{k|k}\}$, $j\in\{1,\dots,i-n_{k|k-1}-1\}$, sent from factor nodes $\underline{s}_k(\widetilde{{\bf X}}_k^i,\beta_k^j;z_k^j)$ to variable nodes $\beta_k^j$. Applying the sum-product rule \cite[Eq. (6)]{kschischang2001factor}, we obtain an integral over the Bernoulli set $\widetilde{{\bf X}}_k^i$
\begin{equation}\label{eq_Theta}
  \Theta_{i,j}^{(p)}\left(\beta_k^j\right) = \int \underline{s}_k\left(\widetilde{{\bf X}}_k^i,\beta_k^j;z_k^j\right)\varepsilon^{(p)}_{i,j}\left(\widetilde{{\bf X}}_k^i\right)\delta \widetilde{{\bf X}}_k^i.
\end{equation}
We then plug \eqref{eq_s_underline} into \eqref{eq_Theta}, which yields
\begin{equation}\label{eq_Theta2}
  \Theta^{(p)}_{i,j}\left(\beta_k^j\right) = \begin{cases}
    \frac{\int \ell_k\left(z_k^j|X\right)\gamma_k(X)\varepsilon^{(p)}_{i,j}\left(\{X\}\right)d X}{\lambda_k^C\left(z_k^j\right)} & \beta_k^j = i\\
    \varepsilon^{(p)}_{i,j} & \beta_k^j \neq i.
  \end{cases}
\end{equation}
After normalizing \eqref{eq_Theta2} by $\varepsilon^{(p)}_{i,j}$, we obtain
\begin{equation}\label{eq_vartheta1}
  \vartheta^{(p)}_{i,j}\left(\beta_k^j\right) = \begin{cases}
    \frac{\int \ell_k\left(z_k^j|X\right)\gamma_k(X)\varepsilon^{(p)}_{i,j}\left(\{X\}\right)d X}{\lambda_k^C\left(z_k^j\right)\varepsilon^{(p)}_{i,j}} & \beta_k^j = i\\
    1 & \beta_k^j \neq i.
  \end{cases}
\end{equation}
Note that multiplying messages by a constant does not alter the resulting approximate marginals \cite{kschischang2001factor}, but the above normalization step makes it possible to perform data association and measurement update more efficiently. 

The messages $\vartheta_{i,j}^{(p)}(\cdot)$, with $i\in\{n_{k|k-1}+1,\dots,n_{k|k}\}$, $j= i-n_{k|k-1}$, sent from factor nodes $\overline{s}_k(\widetilde{{\bf X}}_k^i,\beta_k^j;z_k^j)$ to variable nodes $\beta_k^j$ can be computed similarly. In particular, we have
\begin{equation}\label{eq_vartheta2}
  \vartheta^{(p)}_{i,j}\left(\beta_k^j\right) = \begin{cases}
    \frac{\int \ell_k\left(z_k^j|X\right)\gamma_k(X)\varepsilon^{(p)}_{i,j}\left(\{X\}\right)d X}{\lambda_k^C\left(z_k^j\right)\varepsilon^{(p)}_{i,j}(\emptyset)} & \beta_k^j = i\\
    1 & \beta_k^j \neq i.
  \end{cases}
\end{equation}

\subsubsection{Data association}
We proceed to describe the messages $\varphi_{j,i}^{(p)}(\cdot)$, with $i\in\{1,\dots,n_{k|k-1}\}$, $j\in\{1,\dots,m_k\}$ and $i\in\{n_{k|k-1}+1,\dots,n_{k|k}\}$, $j\in\{1,\dots,i-n_{k|k-1}\}$, sent from variable nodes $\beta_k^j$ to factor nodes $\underline{s}_k(\widetilde{{\bf X}}_k^i,\beta_k^j;z_k^j)$ and $\overline{s}_k(\widetilde{{\bf X}}_k^i,\beta_k^j;z_k^j)$, which can be expressed as \cite[Eq. (5)]{kschischang2001factor}
\begin{equation}\label{eq_varphi}
  \varphi_{j,i}^{(p)}\left(\beta_k^j\right) = \prod_{\substack{i^\prime=1\\i^\prime\neq i}}^{n_{k|k}}\vartheta^{(p)}_{i^\prime,j}\left(\beta_k^j\right).
\end{equation}
By plugging \eqref{eq_vartheta1} and \eqref{eq_vartheta2} into \eqref{eq_varphi}, we obtain 
\begin{equation}\label{eq_varphi2}
  \varphi_{j,i}^{(p)}\left(\beta_k^j\right) = \begin{cases}
    1 & \beta_k^j \in \{0,i\}\\
    \vartheta^{(p)}_{i,j}\left(i\right) & \beta_k^j \in \{1,\dots,n_{k|k}\}\setminus \{i\}.
  \end{cases}
\end{equation}

\subsubsection{Measurement update}
Next, the messages $\varrho_{j,i}^{(p)}(\cdot)$, with $i\in\{1,\dots,n_{k|k-1}\}$, $j\in\{1,\dots,m_k\}$ and $i\in\{n_{k|k-1}+1,\dots,n_{k|k}\}$, $j\in\{1,\dots,i-n_{k|k-1}-1\}$, sent from factor nodes $\underline{s}_k(\widetilde{{\bf X}}_k^i,\beta_k^j;z_k^j)$ to variable nodes $\widetilde{{\bf X}}_k^i$ can be computed as \cite[Eq. (6)]{kschischang2001factor}
\begin{equation}\label{eq_varrho1}
  \Upsilon_{j,i}^{(p)}\left(\widetilde{{\bf X}}_k^i\right) = \sum_{\beta_k^j}\underline{s}_k\left(\widetilde{{\bf X}}_k^i,\beta_k^j;z_k^j\right)\varphi_{j,i}^{(p)}\left(\beta_k^j\right).
\end{equation}
By plugging \eqref{eq_s_underline} and \eqref{eq_varphi2} into \eqref{eq_varrho1}, we obtain
\begin{align}
  &\Upsilon_{j,i}^{(p)}\left(\widetilde{{\bf X}}_k^i\right)\nonumber\\ &= \begin{cases}
    \left(\frac{\ell_k\left(z_k^j|X\right)\gamma_k(X)}{\lambda_k^C\left(z_k^j\right)} + \xi_{i,j}^{(p)}\right)\delta_i[u] & \widetilde{{\bf X}}_k^i = \{(u,X)\}\\
    \xi_{i,j}^{(p)} & \widetilde{{\bf X}}_k^i = \emptyset
  \end{cases}\label{eq_Upsilon3}
\end{align}
where we denote
\begin{equation}\label{eq_xi}
  \xi_{i,j}^{(p)} = \sum_{\substack{i^\prime = 1\\i^\prime \neq i}}^{n_{k|k}}\vartheta^{(p)}_{i^\prime,j}\left(i^\prime\right) +1.
\end{equation}
After normalizing \eqref{eq_Upsilon3} by $\xi_{i,j}^{(p)}$, we obtain
\begin{align}
  &\varrho_{j,i}^{(p)}\left(\widetilde{{\bf X}}_k^i\right)\nonumber\\ &= \begin{cases}
    \left(\frac{\ell_k\left(z_k^j|X\right)\gamma_k(X)}{\lambda_k^C\left(z_k^j\right)\xi_{i,j}^{(p)}} + 1\right)\delta_i[u] & \widetilde{{\bf X}}_k^i = \{(u,X)\}\\
    1 & \widetilde{{\bf X}}_k^i = \emptyset.
  \end{cases}\label{eq_varrho3}
\end{align}

The messages $\varrho_{j,i}^{(p)}(\cdot)$, with $i\in\{n_{k|k-1}+1,\dots,n_{k|k}\}$, $j= i-n_{k|k-1}$, sent from factor nodes $\overline{s}_k(\widetilde{{\bf X}}_k^i,\beta_k^j;z_k^j)$ to variable nodes $\widetilde{{\bf X}}_k^i$ can be  computed similarly. In particular, we have
\begin{equation}
  \varrho_{j,i}^{(p)}\left(\widetilde{{\bf X}}_k^i\right) = \begin{cases}
    \frac{\ell_k\left(z_k^j|X\right)\gamma_k(X)}{\lambda_k^C\left(z_k^j\right)\xi_{i,j}^{(p)}}\delta_i[u] &\widetilde{{\bf X}}_k^i = \{(u,X)\}\\
    1 & \widetilde{{\bf X}}_k^i= \emptyset.
  \end{cases}\label{eq_varrho4}
\end{equation}

\subsubsection{Extrinsic information}
At last, the messages $\varepsilon_{i,j}^{(p+1)}(\cdot)$ at iteration $p+1$, sent from variable nodes $\widetilde{{\bf X}}_k^i$ to factor nodes $\underline{s}_k(\widetilde{{\bf X}}_k^i,\beta_k^j;z_k^j)$, with $i\in\{1,\dots,n_{k|k-1}\}$, $j\in\{1,\dots,m_k\}$, can be computed as \cite[Eq. (5)]{kschischang2001factor}
\begin{equation}\label{eq_extrinsic1}
  \varepsilon_{i,j}^{(p+1)}\left(\widetilde{{\bf X}}_k^i\right) = \underline{f}^{i}_{k|k-1}\left(\widetilde{{\bf X}}_k^i\right)\prod_{\substack{j^\prime=1\\j^\prime\neq j}}^{m_k}\varrho_{j^\prime,i}^{(p)}\left(\widetilde{{\bf X}}_k^i\right),
\end{equation}
and similarly, for the messages $\varepsilon_{i,j}^{(p+1)}(\cdot)$, sent from variable nodes $\widetilde{{\bf X}}_k^i$ to factor nodes $\underline{s}_k(\widetilde{{\bf X}}_k^i,\beta_k^j;z_k^j)$ or $\overline{s}_k(\widetilde{{\bf X}}_k^i,\beta_k^j;z_k^j)$, with $i\in\{n_{k|k-1}+1,\dots,n_{k|k}\}$, $j\in\{1,\dots,i-n_{k|k-1}\}$, we have
\begin{equation}\label{eq_extrinsic2}
  \varepsilon_{i,j}^{(p+1)}\left(\widetilde{{\bf X}}_k^i\right) = \overline{f}_{k|k-1}\left(\widetilde{{\bf X}}_k^i\right) \prod_{\substack{j^\prime=1\\j^\prime\neq j}}^{i-n_{k|k-1}}\varrho_{j^\prime,i}^{(p)}\left(\widetilde{{\bf X}}_k^i\right).
\end{equation}

\subsection{Belief calculation}
After the last iteration $p=P$, we can evaluate the belief $\widetilde{f}^i_{k|k}(\cdot)$ for each Bernoulli component $i\in\{1,\dots,n_{k|k}\}$, which is proportional to the product of all coming messages \cite{kschischang2001factor}. In particular, for $i\in\{1,\dots,n_{k|k-1}\}$,
\begin{equation}
  \widetilde{f}^i_{k|k}\left(\widetilde{{\bf X}}_k^i\right)\propto \underline{f}^{i}_{k|k-1}\left(\widetilde{{\bf X}}_k^i\right)\prod_{\substack{j=1}}^{m_k}\varrho_{j,i}^{(P)}\left(\widetilde{{\bf X}}_k^i\right),
\end{equation}
and for $i\in\{n_{k|k-1}+1,\dots,n_{k|k}\}$,
\begin{equation}
  \widetilde{f}^i_{k|k}\left(\widetilde{{\bf X}}_k^i\right)\propto \overline{f}^i_{k|k-1}\left(\widetilde{{\bf X}}_k^i\right) \prod_{\substack{j=1}}^{i-n_{k|k-1}}\varrho_{j,i}^{(P)}\left(\widetilde{{\bf X}}_k^i\right)
\end{equation}
where normalization is required to ensure that $\widetilde{f}^i_{k|k}(\cdot)$ is a valid set density. 

\section{Particle Implementation of Trajectory PMB Filters Using Belief Propagation}

For general multi-object dynamic and measurement models, the messages in BP typically cannot be evaluated in closed form. In this section, we present the particle-based implementation\footnote{In theory, it is possible to use a Rao-Blackwellized particle filter for BP, where we only sample kinematic states while keeping an analytic representation of object extent state conditioned on its kinematic state. However, because of the way extrinsic information \eqref{eq_extrinsic1}, \eqref{eq_extrinsic2} is computed, doing so requires an analytic representation for every possible non-empty subset of the measurements, resulting in an exponential growth of local hypotheses.} of the TPMB filter with Poisson spatial measurement model for both the set of alive trajectories and the set of all trajectories. In addition, we discuss aspects that need to be considered in practical implementations and how the proposed implementation compares to the one in \cite{florian2021scalable}. We note that only the update step of the proposed TPMB filters involves BP.

We represent a single-trajectory density/intensity by
\begin{equation*}
  f\left(t,x^{1:\nu}\right) = \sum_{l=1}^Lw^{(l)}\delta_{t^{(l)}}[t]\delta_{\nu^{(l)}}[\nu]\delta_{\chi^{(l)}}\left(x^{1:\nu}\right)
\end{equation*}
where the $l$-th particle has weight $w^{(l)}\geq 0$, and it represents a single trajectory with start time $t^{(l)}$ and sequence of object states $\chi^{(l)}\in {\cal X}^{\nu^{(l)}}$, where the length $\nu^{(l)}$ is implicit in $\chi^{(l)}$ \cite{xia2022multiple}. In addition, we denote the last state of $\chi^{(l)}$ as $\overline{\chi}^{(l)}$. Furthermore, we note that $f(\cdot)$ is a density if and only if $\sum_{l=1}^Lw^{(l)} = 1$, and that it can be fully described by the set of parameters $\{(w^{(l)},t^{(l)},\chi^{(l)})\}_{l=1}^L$.

For the set of alive trajectories, the single-trajectory density of the $i$-th Bernoulli component is of the form
\begin{equation}\label{eq_particle_tra_ber}
  f^i_{k|k^\prime}\left(t,x^{1:\nu}\right) = \sum_{l=1}^{L^d_{k|k^\prime}}w^{i,(l)}_{k|k^\prime}\delta_{t_{k|k^\prime}^{i,(l)}}[t]\delta_{k-t_{k|k^\prime}^{i,(l)}+1}[\nu]\delta_{\chi_{k|k^\prime}^{i,(l)}}\left(x^{1:\nu}\right),
\end{equation}
which implies that, if the corresponding trajectory exists, it is alive at time step $k$ with probability one. The PPP for undetected trajectories has intensity
\begin{equation}\label{eq_particle_tra_ppp}
  \lambda_{k|k^\prime}\left(t,x^{1:\nu}\right) =  \sum_{l=1}^{L^u_{k|k^\prime}}w^{0,(l)}_{k|k^\prime}\delta_{t_{k|k^\prime}^{0,(l)}}[t]\delta_{k-t_{k|k^\prime}^{0,(l)}+1}[\nu]\delta_{\chi_{k|k^\prime}^{0,(l)}}\left(x^{1:\nu}\right)
\end{equation} 
where $\sum_{l=1}^{L^u_{k|k^\prime}}w^{0,(l)}_{k|k^\prime}$ gives the expected number of undetected trajectories.

For the set of all trajectories, the single-trajectory density of the $i$-th Bernoulli component is of the form 
\begin{equation}\label{eq_particle_tra_ber2}
  f^i_{k|k^\prime}\left(t,x^{1:\nu}\right) = \sum_{l=1}^{L^d_{k|k^\prime}}w^{i,(l)}_{k|k^\prime}\delta_{t_{k|k^\prime}^{i,(l)}}[t]\delta_{\nu_{k|k^\prime}^{i,(l)}}[\nu]\delta_{\chi_{k|k^\prime}^{i,(l)}}\left(x^{1:\nu}\right),
\end{equation}
and the Poisson intensity for undetected trajectories is of the same form as \eqref{eq_particle_tra_ppp}. In the implementation, we do not account for undetected trajectories that are not present at the current time step since these are usually not of practical interest.

\subsection{Prediction step}

\subsubsection{Set of alive trajectories}\label{sec_alive}
Assume that the filtering density for the alive trajectories is a PMB of the form \eqref{eq_pmb_original} with $f^i_{k-1|k-1}(\cdot)$ and $\lambda_{k-1|k-1}(\cdot)$ given by \eqref{eq_particle_tra_ber} and \eqref{eq_particle_tra_ppp}. Then, the predicted density is a PMB of the form \eqref{eq_pmb_original} with 
\begin{subequations}\label{eq_predict_lambda}
  \begin{align}
    \lambda_{k|k-1}\left(t,x^{1:\nu}\right)&= \sum_{l=1}^{L^u_{k-1|k-1}}w^{0,(l)}_{k|k-1}\delta_{t_{k|k-1}^{0,(l)}}[t]\nonumber\\ &~~~\times\delta_{k-t_{k|k-1}^{0,(l)}+1}[\nu]\delta_{\chi_{k|k-1}^{0,(l)}}\left(x^{1:\nu}\right) \nonumber\\ &~~~+ \sum_{l=1}^{L^b}w_k^{b,(l)}\delta_k[t]\delta_1[\nu]\delta_{\chi^{b,(l)}_k}\left(x^{1:\nu}\right),\\
    L^u_{k|k-1} &= L^u_{k-1|k-1} + L^b,\\
    t_{k|k-1}^{0,(l)} &= t_{k-1|k-1}^{0,(l)},\\
    x_k^{0,(l)} &\sim g_{k}\left(\cdot|\overline{\chi}^{0,(l)}_{k-1|k-1}\right),\\
    \chi^{0,(l)}_{k|k-1} &= \left(\chi^{0,(l)}_{k-1|k-1},x_k^{0,(l)}\right),\\
    w^{0,(l)}_{k|k-1} &= p^S\left(\overline{\chi}^{0,(l)}_{k-1|k-1}\right)w^{0,(l)}_{k-1|k-1},\\
    w_k^{b,(l)} &= \left\langle \lambda^B_k,1 \right\rangle/L^b,\\
    \chi^{b,(l)}_k &\sim \lambda_k^B(\cdot)/\left\langle \lambda^B_k,1 \right\rangle
  \end{align}
\end{subequations}
where $L^b$ is the number of particles used to represent newborn trajectories, and we can see that the prediction of undetected trajectories comprises the prediction of existing particles and the generation of new particles. As for the predicted Bernoulli component, it is parameterized by
\begin{subequations}\label{eq_predict_f}
  \begin{align}
    L^d_{k|k-1} &= L^d_{k-1|k-1},\\
    r^i_{k|k-1} &= r^i_{k-1|k-1}\sum_{l=1}^{L^d_{k|k-1}}p^S\left(\overline{\chi}_{k-1|k-1}^{i,(l)}\right)w^{i,(l)}_{k-1|k-1},\\
    t_{k|k-1}^{i,(l)} &= t_{k-1|k-1}^{i,(l)},\\
    x_k^{i,(l)} &\sim g_{k}\left(\cdot|\overline{\chi}^{i,(l)}_{k-1|k-1}\right),\\
    \chi_{k|k-1}^{i,(l)} &= \left(\chi_{k-1|k-1}^{i,(l)},x_k^{i,(l)}\right),\\
    w^{i,(l)}_{k|k-1} &= \frac{p^S\left(\overline{\chi}_{k-1|k-1}^{i,(l)}\right)w^{i,(l)}_{k-1|k-1}}{\sum_{l=1}^{L^d_{k|k-1}}p^S\left(\overline{\chi}_{k-1|k-1}^{i,(l)}\right)w^{i,(l)}_{k-1|k-1}}.
  \end{align}
\end{subequations}

The proposal densities used in the above prediction step for drawing samples are given by the motion model $g_k(\cdot|\cdot)$ and the birth density $\lambda_k^B(\cdot)/\left\langle \lambda^B_k,1 \right\rangle$. If these densities are difficult to sample from, then other proposal densities may be used for importance sampling \cite{sarkka2013bayesian}.

\subsubsection{Set of all trajectories}\label{sec_all}
Assume that the filtering density for the set of all trajectories is a PMB of the form \eqref{eq_pmb_original} with $f^i_{k-1|k-1}(\cdot)$ and $\lambda_{k-1|k-1}(\cdot)$ given by \eqref{eq_particle_tra_ber2} and \eqref{eq_particle_tra_ppp}. Then, the predicted density is a PMB of the form \eqref{eq_pmb_original} with $\lambda_{k|k-1}(\cdot)$ given by \eqref{eq_predict_lambda} and $f^i_{k|k-1}(\cdot)$ parameterized by
\begin{subequations}
  \begin{align}
    r^i_{k|k-1} &= r^i_{k-1|k-1},\\
    L^d_{k|k-1} &= L^d_{k-1|k-1}\nonumber
    \\ &~~~+ \sum_{l=1}^{L^d_{k-1|k-1}}\delta_{k-t^{i,(l)}_{k-1|k-1}}\left[\nu_{k-1|k-1}^{i,(l)}\right],\\
    f^i_{k|k-1}\left(t,x^{1:\nu}\right) &= \sum_{l=1}^{L^d_{k-1|k-1}}w^{i,(l)}_{k|k-1}\delta_{t_{k|k-1}^{i,(l)}}[t]\nonumber\\ &~~~\times \delta_{\nu_{k|k-1}^{i,(l)}}[\nu]\delta_{\chi_{k|k-1}^{i,(l)}}\left(x^{1:\nu}\right)\nonumber\\
    &~~~+\sum_{l=1}^{L^d_{k-1|k-1}}\left[1- p^S\left(\overline{\chi}^{i,(l)}_{k-1|k-1}\right)\right]\nonumber\\
    &~~~\times w^{i,(l)}_{k-1|k-1}\delta_{t_{k-1|k-1}^{i,(l)}}[t]\nonumber\\
    &~~~\times \delta_{k-t^{i,(l)}_{k-1|k-1}}\left[\nu_{k-1|k-1}^{i,(l)}\right]\delta_{\chi_{k-1|k-1}^{i,(l)}}\left(x^{1:\nu}\right),\\
    t_{k|k-1}^{i,(l)} &= t_{k-1|k-1}^{i,(l)},
  \end{align}
  and if $\nu_{k-1|k-1}^{i,(l)} = k-t^{i,(l)}_{k-1|k-1}$,
  \begin{align}
    x_k^{i,(l)} &\sim g_{k}\left(\cdot|\overline{\chi}^{i,(l)}_{k-1|k-1}\right),\\
    \chi_{k|k-1}^{i,(l)} &= \left(\chi_{k-1|k-1}^{i,(l)},x_k^{i,(l)}\right),\\
    w^{i,(l)}_{k|k-1} &= p^S\left(\overline{\chi}_{k-1|k-1}^{i,(l)}\right)w^{i,(l)}_{k-1|k-1},
  \end{align}
  otherwise
  \begin{align}
    \chi_{k|k-1}^{i,(l)} &= \chi_{k-1|k-1}^{i,(l)},\\
    w^{i,(l)}_{k|k-1} &= w^{i,(l)}_{k-1|k-1}.
  \end{align}
\end{subequations}

We note that for each particle used to represent a Bernoulli component, it remains unchanged if it corresponds to a dead trajectory, and that a copy of it is created if it corresponds to an alive trajectory. In the latter case, there is a change in the weight for original particles to account for the probability that the object dies, and for each copy, its weight is updated with the survival probability and its state is propagated to the next time step. The dimensions of particles may be high for long trajectories, but this does not mean that the computation would be infeasible. In the standard particle filter implementations, the particle trajectory length also increases in time, see, e.g., \cite[Eq. (40)]{arulampalam2002tutorial}. The particle degeneracy problem in practical implementations is dealt in Section \ref{sec_estimation}. We also note that the particle implementation of the TPMB prediction step is general, which also holds for TPMBM and is not limited to EOT.

\subsection{Update step}
The update step is the same for both the set of alive trajectories and the set of all trajectories, where for each Bernoulli component we only update its probability of existence and the weights of particles while leaving the states of particles unchanged. Assume that the predicted density is a PMB of the form \eqref{eq_pmb_original} with parameters described in either Section \ref{sec_alive} or Section \ref{sec_all}, the updated Poisson intensity $\lambda_{k|k}(\cdot)$ for undetected trajectories is described by
\begin{equation*}
  \left\{ \left(t^{0,(l)}_{k|k-1},\chi^{0,(l)}_{k|k-1} w_{k|k-1}^{0,(l)}e^{-\gamma_k\left(\overline{\chi}_{k|k-1}^{0,(l)}\right)} \right) \right\}_{l=1}^{L^u_{k|k-1}},
\end{equation*} 
and the implementation of particle BP is given as follows, where we omit the auxiliary variables in all the messages and beliefs for notational brevity.

\subsubsection{Initialization}\label{sec_initial}
Each message $\varepsilon_{i,j}^{(p)}(\cdot)$ is represented by a scalar and a set of weighted particles 
\begin{equation*}
  \left(\varepsilon_{i,j}^{(p)}, \left\{\left( t_{k|k}^{i,(l)},\chi_{k|k}^{i,(l)},w_{i,j}^{(p,l)} \right) \right\}_{l=1}^{L_i}\right).
\end{equation*}
For $p=1$ and $i\in\{1,\dots,n_{k|k-1}\}$, we set 
\begin{align}
  &\left\{\left(t_{k|k}^{i,(l)},\chi_{k|k}^{i,(l)}\right)\right\}_{l=1}^{L_i} = \left\{ \left(t^{i,(l)}_{k|k-1},\chi_{k|k-1}^{i,(l)}\right)\right\}_{l=1}^{L^d_{k|k-1}},\nonumber\\
  w_{i,j}^{(1,l)} &= \begin{cases}
    w^{i,(l)}_{k|k-1} e^{-\gamma_k\left(\overline{\chi}_{k|k-1}^{i,(l)}\right)} & k - t^{i,(l)}_{k|k-1}+1 = \nu^{i,(l)}_{k|k-1}\\
    w^{i,(l)}_{k|k-1} & \text{otherwise},
  \end{cases}\nonumber\\
  \varepsilon_{i,j}^{(1)} &= r^i_{k|k-1}\sum_{l=1}^{L^d_{k|k-1}}\delta_{k-t^{i,(l)}_{k|k-1}+1}\left[\nu_{k|k-1}^{i,(l)}\right]w_{k|k-1}^{i,(l)} \nonumber\\ &~~~\times e^{-\gamma_k\left(\overline{\chi}_{k|k-1}^{i,(l)}\right)} +  1 - r^i_{k|k-1},\nonumber
\end{align}
and for $i\in\{n_{k|k-1}+1,\dots,n_{k|k}\}$, we set
\begin{align}
  &\left\{\left(t_{k|k}^{i,(l)},\chi_{k|k}^{i,(l)},w_{i,j}^{(1,l)}\right) \right\}_{l=1}^{L_i}\nonumber\\ &= \left\{ \left(t^{0,(l)}_{k|k-1},\chi_{k|k-1}^{0,(l)},w_{k|k-1}^{0,(l)}e^{-\gamma_k\left(\overline{\chi}_{k|k-1}^{0,(l)}\right)} \right) \right\}_{l=1}^{L^u_{k|k-1}},\nonumber\\
  \varepsilon_{i,j}^{(1)} &= \sum_{l=1}^{L^u_{k|k-1}}w^{0,(l)}_{k|k-1}e^{-\gamma_k\left(\overline{\chi}_{k|k-1}^{0,(l)}\right)} + 1.\nonumber
\end{align}

\subsubsection{Measurement evaluation}
For $i\in\{1,\dots,n_{k|k-1}\}$, $j\in\{1,\dots,m_k\}$ and $i\in\{n_{k|k-1}+1,\dots,n_{k|k}\}$, $j\in\{1,\dots,i-n_{k|k-1}-1\}$, we have 
\begin{align}\label{eq_meas_eva_particle}
  \vartheta^{(p)}_{i,j}\left(i\right) &= \frac{1}{\lambda^C_k\left(z_k^j\right)\varepsilon_{i,j}^{(p)}}\sum_{l=1}^{L_i}w_{i,j}^{(p,l)}\delta_{k-t^{i,(l)}_{k|k}+1}\left[\nu_{k|k}^{i,(l)}\right]\nonumber\\
  &~~~\times \gamma_k\left(\overline{\chi}_{k|k}^{i,(l)}\right)\ell_k\left(z_k^j|\overline{\chi}_{k|k}^{i,(l)}\right)
\end{align}
As for $i\in\{n_{k|k-1}+1,\dots,n_{k|k}\}$, $j= i-n_{k|k-1}$, the message $\vartheta^{(p)}_{i,j}(i)$ is obtained by replacing $\varepsilon_{i,j}^{(p)}$ with $\varepsilon_{i,j}^{(p)}(\emptyset)$ in \eqref{eq_meas_eva_particle}. It should be noted that in \eqref{eq_meas_eva_particle} we only consider particles that represent alive trajectories.

\subsubsection{Data association, measurement update, and extrinsic information}
The data association and measurement update steps do not need to be explicitly implemented. After computing $\vartheta^{(p)}_{i,j}(\cdot)$, we can then obtain $\xi_{j,i}^{(p)}(\cdot)$ using \eqref{eq_xi} and \eqref{eq_meas_eva_particle}. Each message $\varepsilon_{i,j}^{(p+1)}(\cdot)$ is an unnormalized Bernoulli density, represented by a scalar and a set of weighted particles 
\begin{equation*}
  \left(\varepsilon_{i,j}^{(p+1)}, \left\{\left(t_{k|k}^{i,(l)},\chi_{k|k}^{i,(l)},w_{i,j}^{(p+1,l)} \right)\right\}_{l=1}^{L_i}\right),
\end{equation*}
which can be calculated by first computing the particle-based representation of \eqref{eq_varrho3}/\eqref{eq_varrho4} and then plugging it into \eqref{eq_extrinsic1}/\eqref{eq_extrinsic2}. This yields, for $i\in\{1,\dots,n_{k|k-1}\}$, $j\in\{1,\dots,m_k\}$,
\begin{align}
  \varepsilon_{i,j}^{(p+1)} &= r^i_{k|k-1}\sum_{l=1}^{L_i}w_{i,j}^{(p+1,l)} + 1-r^i_{k|k-1},\\
  w_{i,j}^{(p+1,l)} &= w_{i,j}^{(1,l)}\prod_{\substack{j^\prime=1\\j^\prime\neq j}}^{m_k}\Bigg[\frac{\ell_k\left(z_k^{j^\prime}|\overline{\chi}_{k|k}^{i,(l)}\right)\gamma_k\left(\overline{\chi}_{k|k}^{i,(l)}\right)}{\lambda_k^C\left(z_k^{j^\prime}\right)\xi_{i,j^\prime}^{(p)}}\nonumber\\
  &~~~\times \delta_{k-t^{i,(l)}_{k|k}+1}\left[\nu_{k|k}^{i,(l)}\right] + 1\Bigg],\label{eq_particle_extrinsic1}
\end{align}
for $i\in\{n_{k|k-1}+1,\dots,n_{k|k}\}$, $j\in\{1,\dots,i-n_{k|k-1}\}$,
\begin{equation}
  \varepsilon_{i,j}^{(p+1)} = \sum_{l=1}^{L_i}w_{i,j}^{(p+1,l)} + 1,
\end{equation}
for $i\in\{n_{k|k-1}+1,\dots,n_{k|k}\}$, $j\in\{1,\dots,i-n_{k|k-1}-1\}$,
\begin{align}
  w_{i,j}^{(p+1,l)} &= w_{i,j}^{(1,l)}\frac{\ell_k\left(z_k^{i-n_{k|k-1}}|\overline{\chi}_{k|k}^{i,(l)}\right)\gamma_k\left(\overline{\chi}_{k|k}^{i,(l)}\right)}{\lambda_k^C\left(z_k^{i-n_{k|k-1}}\right)\xi_{i,i-n_{k|k-1}}^{(p)}}\nonumber\\
  &~~~\times\prod_{\substack{j^\prime=1\\j^\prime\neq j}}^{i-n_{k|k-1}-1}\left[\frac{\ell_k\left(z_k^{j^\prime}|\overline{\chi}_{k|k}^{i,(l)}\right)\gamma_k\left(\overline{\chi}_{k|k}^{i,(l)}\right)}{\lambda_k^C\left(z_k^{j^\prime}\right)\xi_{i,j^\prime}^{(p)}} + 1\right],\label{eq_particle_extrinsic2}
\end{align}
and for $i\in\{n_{k|k-1}+1,\dots,n_{k|k}\}$, $j=i-n_{k|k-1}$,
\begin{equation}
  w_{i,j}^{(p+1,l)} = w_{i,j}^{(1,l)}\prod_{\substack{j^\prime=1}}^{j-1}\left[\frac{\ell_k\left(z_k^{j^\prime}|\overline{\chi}_{k|k}^{i,(l)}\right)\gamma_k\left(\overline{\chi}_{k|k}^{i,(l)}\right)}{\lambda_k^C\left(z_k^{j^\prime}\right)\xi_{i,j^\prime}^{(p)}} + 1\right].
\end{equation}

\subsubsection{Belief calculation}
Each belief $\widetilde{f}^i_{k|k}(\cdot)$ is a Bernoulli RFS density, parameterized by 
\begin{equation*}
  \left(r^i_{k|k},\left\{\left(t^{i,(l)}_{k|k},\chi^{i,(l)}_{k|k},w^{i,(l)}_{k|k}\right)\right\}_{l=1}^{L_i}\right),
\end{equation*}
which can be calculated similarly as the extrinsic information $\varepsilon_{i,j}^{(p+1)}(\cdot)$ with the difference that the product in \eqref{eq_particle_extrinsic1} and \eqref{eq_particle_extrinsic2} now needs to enumerate every element, and that normalization is required to ensure that $\widetilde{f}^i_{k|k}(\cdot)$ is a valid density.

Specifically, we have, for $i\in\{1,\dots,n_{k|k-1}\}$,
\begin{align}
  w_i^{(P,l)} &= w_{i,1}^{(1,l)}\prod_{\substack{j^\prime=1}}^{m_k}\Bigg[\frac{\ell_k\left(z_k^{j^\prime}|\overline{\chi}_{k|k}^{i,(l)}\right)\gamma_k\left(\overline{\chi}_{k|k}^{i,(l)}\right)}{\lambda_k^C\left(z_k^{j^\prime}\right)\xi_{i,j^\prime}^{(P)}}\nonumber\\
  &~~~\times \delta_{k-t^{i,(l)}_{k|k}+1}\left[\nu_{k|k}^{i,(l)}\right] + 1\Bigg],\\
  r^i_{k|k} &= \frac{r^i_{k|k-1}\sum_{l=1}^{L_i}w_i^{(P,l)}}{r^i_{k|k-1}\sum_{l=1}^{L_i}w_i^{(P,l)}+1-r^i_{k|k-1}},
\end{align}
and for $i\in\{n_{k|k-1}+1,\dots,n_{k|k}\}$,
\begin{align}
  w_i^{(P,l)} &= w_{i,1}^{(1,l)}\frac{\ell_k\left(z_k^{i-n_{k|k-1}}|\overline{\chi}_{k|k}^{i,(l)}\right)\gamma_k\left(\overline{\chi}_{k|k}^{i,(l)}\right)}{\lambda_k^C\left(z_k^{i-n_{k|k-1}}\right)\xi_{i,i-n_{k|k-1}}^{(P)}}\nonumber\\
  &~~~\times\prod_{\substack{j^\prime=1}}^{i-n_{k|k-1}-1}\left[\frac{\ell_k\left(z_k^{j^\prime}|\overline{\chi}_{k|k}^{i,(l)}\right)\gamma_k\left(\overline{\chi}_{k|k}^{i,(l)}\right)}{\lambda_k^C\left(z_k^{j^\prime}\right)\xi_{i,j^\prime}^{(P)}} + 1\right],\\
  r^i_{k|k} &= \frac{\sum_{l=1}^{L_i}w_i^{(P,l)}}{\sum_{l=1}^{L_i}w_i^{(P,l)}+1},
\end{align}
and the normalized weight is given by
\begin{equation}
  w^{i,(l)}_{k|k} = \frac{w_i^{(P,l)}}{\sum_{l=1}^{L_i}w_i^{(P,l)}}.
\end{equation}

\subsection{Practical considerations}

\subsubsection{Approximations for efficient implementation}
For the proposed particle implementation of TPMB, the computational complexity of its prediction step scales linearly in the number of Bernoulli components (potential trajectories). As for the update step, its computational complexity for a fixed number of message passing iterations scales quadratically in the number of Bernoulli components and measurements \cite{florian2021scalable}.

Since the number of Bernoulli components increases with time, we need to prune Bernoulli components with probability of existence smaller than a threshold. For the implementation considering the set of all trajectories, the number of particles in the Bernoulli components quickly increases over time as more particles are used to represent dead trajectories in the prediction step. To avoid this, for the $i$-th Bernoulli component, $i\in\{1,\dots,n_{k|k-1}\}$, we first compute the probability mass function of the trajectory end time $t^i_e$ 
\begin{equation}
  P\left(t^{i}_{e}=t\right) = \sum_{l=1}^{L^d_{k|k}}w^{i,(l)}_{k|k}\delta_{t}\left[t^{i,(l)}_{k|k} + \nu_{k|k}^{i,(l)} - 1\right],\label{eq_card_death}
\end{equation}
and find those $t^i_e$ with probability \eqref{eq_card_death} smaller than a threshold. Then, we discard particles with the corresponding trajectory end times and re-normalize the particle weights and the probability of existence $r^i_{k|k}$.  At last, we note that computational complexity can be further reduced by censoring of messages and reordering of measurements, see \cite{florian2020scalable} for details.

\subsubsection{Measurement-driven initialization of newly detected trajectories}\label{sec_measure_driven}
In the initialization step of particle BP described in Section \ref{sec_initial}, particles describing newly detected trajectories are set as particles representing undetected trajectories. For non-informative birth densities, e.g., a single Gaussian with very large covariance and uniform distribution, it becomes more advantageous to use importance sampling and directly draw samples from a proposal density related to the measurement likelihood $\ell_k(\cdot|\cdot)$\footnote{If $\ell_k(z_k|x_k)$ does not depend on all the elements of $x_k$, one can draw samples of elements of $x_k$ that do not depend on $z_k$ from a suitable proposal density, e.g., a Gaussian or a uniform distribution.} \cite{kropfreiter2016sequential}.

Specifically, in the initialization step, for $i\in\{n_{k|k-1}+1,\dots,n_{k|k}\}$ and $l\in\{1,\dots,L^u_{k|k-1}\}$, we draw sample $\chi_{k|k}^{i,(l)}$ from a proposal density $\ell^\prime_k(\cdot|z_{i-n_{k|k-1}})$, and its corresponding weight $w_{i,j}^{(1,l)}$ is given by
\begin{equation*}
  w_{i,j}^{(1,l)} \propto \frac{\sum_{l=1}^{L^u_{k|k-1}} w^{0,(l)}_{k|k-1}e^{-\gamma_k\left(\overline{\chi}_{k|k-1}^{0,(l)}\right)} K(\chi_{k|k}^{i,(l)}-\overline{\chi}_{k|k-1}^{0,(l)}) }{\ell^\prime_k(\chi_{k|k}^{i,(l)}|z_{i-n_{k|k-1}})}
\end{equation*}
where $K(\cdot)$ is a user-defined kernel function \cite{kropfreiter2016sequential}, typically a multivariate Gaussian with suitable covariance. At last, the weight $w_{i,j}^{(1,l)}$ needs to be normalized such that its sum is given by $\sum_{l=1}^{L^u_{k|k-1}}w^{0,(l)}_{k|k-1}e^{-\gamma_k\left(\overline{\chi}_{k|k-1}^{0,(l)}\right)}$.

In practical scenarios, the birth density is typically uniform or a Gaussian mixture. For these cases with state-independent Poisson measurement rate $\gamma_k$, it is not necessary to consider particle representation of the Poisson intensity of undetected trajectories $\lambda_{k|k^\prime}(\cdot)$, and consequently there is no need to draw particles in the prediction step for undetected trajectories. In particular, when the birth density is uniform, $\lambda_{k|k^\prime}(\cdot)$ can be approximately represented using only a scalar \cite{kropfreiter2016sequential}, whereas when the birth density is a Gaussian mixture, $\lambda_{k|k^\prime}(\cdot)$ can be computed using the Gaussian implementation in \cite{garcia2020trajectory}. For the more general cases, one can also use a grid of points to model the Poisson intensity of undetected objects \cite{bostrom2021pmbm}.

\subsubsection{Trajectory estimation}\label{sec_estimation}
One problem with sequential importance sampling is weight degeneracy, which can be reduced by resampling particles that represent alive trajectories \cite{sarkka2013bayesian}. Furthermore, resampling can be used to cap the number of particles in single-trajectory density/intensity representation. We also note that, by keeping the full sample histories in the prediction step, particle filtering provides an approximation to the smoothing problem as a by-product. However, the resulting approximation also tends to be degenerate for alive trajectories at some point in the past as many particles may share the same history at earlier time steps \cite{saha2012particle}. We proceed to describe how to obtain reasonable trajectory estimates in the proposed TPMB implementation.

We first select Bernoulli components with existence probability above a threshold, and then from each of them, we obtain a single-trajectory estimate from a set of weighted particles (i.e., a mixture of Dirac delta single-trajectory densities) as follows. At time step $k^\prime$, given the particle representation of single-trajectory density of the $i$-th Bernoulli component,
\begin{equation*}
  \left\{\left(w_{k^\prime|k^\prime}^{i,(l)},t_{k^\prime|k^\prime}^{i,(l)},\chi_{k^\prime|k^\prime}^{i,(l)}\right)\right\}_{l=1}^{L^d_{k^\prime|k^\prime}},
\end{equation*}
we extract and store the particle representation of the marginal single-object state at time step $k^\prime$,
\begin{subequations}\label{eq_storedparticles}
  \begin{align}
    &\left\{\left(w_{k^\prime|k^\prime}^{i,(l)},\overline{\chi}_{k^\prime|k^\prime}^{i,(l)}\right): l \in \mathbb{L}^i_{k^\prime} \right\},\\
    &\mathbb{L}^{i}_{k^\prime} = \left\{l:k^\prime - t_{k^\prime|k^\prime}^{i,(l)} + 1 = \nu_{k^\prime|k^\prime}^{i,(l)}\right\}
  \end{align}
\end{subequations}
where $\nu_{k^\prime|k^\prime}^{i,(l)}$ is the length of ${\chi}_{k^\prime|k^\prime}^{i,(l)}$. To extract the trajectory estimate at time step $k^\prime \leq k$, we first compute the probability mass functions of the trajectory start time $t^i_s$ and end time $t^i_e$, which are given by
\begin{equation}
  P\left(t^{i}_s=t\right) = \sum_{l=1}^{L^d_{k|k}}w^{i,(l)}_{k|k}\delta_{t}\left[t^{i,(l)}_{k|k}\right],
\end{equation}
and \eqref{eq_card_death}, respectively. Note that for the implementation considering the set of current trajectories, we have $P(t^i_e = k) = 1$.

Next, we find the maximum a posteriori estimates $\hat{t}^i_s$ and $\hat{t}^i_e$ of the trajectory start and end times. Then, we can obtain the object state estimate at time step $k^\prime$ with $\hat{t}^i_s \leq k^\prime \leq \hat{t}^i_e$ by
\begin{equation}
  \hat{x}^i_{k^\prime} = \frac{1}{\sum_{l\in \mathbb{L}_{k^\prime}^i}w^{i,(l)}_{k^\prime|k^\prime}}\sum_{l\in{\mathbb{L}_{k^\prime}^i}}w^{i,(l)}_{k^\prime|k^\prime}\overline{\chi}_{k^\prime|k^\prime}^{i,(l)},
\end{equation}
where $\{(w_{k^\prime|k^\prime}^{i,(l)},\overline{\chi}_{k^\prime|k^\prime}^{i,(l)}): l \in \mathbb{L}^i_{k^\prime} \}$ are pre-stored particles \eqref{eq_storedparticles} at previous time steps. This can be understood as using a type of $L$-scan approximation \cite{garcia2020trajectory} with $L=1$ for the object states. In the general $L$-scan approximation \cite{garcia2019trajectory}, past states of the trajectories before the last $L$ time steps are considered independent, and no smoothing is performed when $L=1$. For Gaussian implementations, a larger $L$ yields more smoothed trajectory estimates. As for particle implementations, using a large $L$ does not necessarily improve the trajectory estimates due to the particle history degeneracy problem. 

At last, we note that, if desired, individual smoothed trajectory estimate can be obtained using, e.g., backward simulation \cite{lindsten2013backward}, which is described in Appendix \ref{appendix_d}. Backward simulation can be applied either only at the final time step or in a sliding window in the form of fixed-lag smoothing. Doing so does not improve the accuracy of the data association results or the object cardinality estimates. It is also possible to consider joint backward simulation of the whole multi-trajectory distribution \cite{xia2022multiple}, where improved data associations can be obtained during backward smoothing.

\subsection{Discussion}

The connection between the proposed TPMB filter and the SPA filter \cite{florian2021scalable} can be understood as follows. The SPA filter may be seen as a TPMB filter for the set of alive trajectories, where previous object states are marginalized out and the PPP intensity of undetected objects is set to zero after the update step. From another perspective, the TPMB filter may be regarded as a derivation of the SPA filter using RFSs, with undetected objects/trajectories propagated over time in parallel 
using the standard PMBM equations for the PPP, which are equivalent to a zero-measurement PHD filter \cite{Horridge2011UsingAP}.

There are several important differences between the proposed TPMB filters and the SPA filter in \cite{florian2021scalable}. First, the set of undetected objects is explicitly modelled as a PPP in the TPMB filters, and its Poisson intensity is propagated over time. This becomes advantageous when the state-dependent Poisson measurement rate $\gamma_k(\cdot)$ is small, which could happen in, e.g., scenarios with low sensor resolution, occlusion, or long distance between the sensor and the objects. One such application is EOT with long-range automotive radar. Second, the object survival probability $p^S(\cdot)$ is state-dependent in the TPMB filters.

Lastly, the TPMB filter directly reports trajectory estimates from local trajectory hypotheses (trajectory Bernoulli components), and therefore these trajectory estimates do not have any gaps in between. As a comparison, the SPA filter in \cite{florian2021scalable} only estimates the set of object states at the current time step, and post-processing is required to link object state estimates at different time steps and to bridge possible gaps. In addition, although for both implementations smoothed object state estimates can be obtained using particle smoothing techniques, the particle TPMB filters also provide smoothed estimate of the start time and length of trajectories, in the sense that estimates at early time steps may be refined at later time steps. This is an important feature for MOT algorithms based on sets of trajectories.

\section{Simulation Results}

In this section, we present the results from a Monte Carlo simulation with 200 runs where the performance of the following multiple extended object trackers are compared\footnote{MATLAB implementations of TPMB-BP and PMB-BP are available at \url{https://github.com/yuhsuansia/Trajectory-PMB-EOT-BP}. MATLAB implementation of SPA is available at \url{https://github.com/meyer-ucsd/EOT-TSP-21}. MATLAB implementations of TPMB(M)-CA and TPMB(M)-SO are available at \url{https://github.com/yuhsuansia/Extended-target-PMBM-tracker}.}:
\begin{itemize}
  \item Trajectory PMB filter using BP, referred to as TPMB-BP.
  \item PMB filter using BP, referred to as PMB-BP.
  \item The SPA filter \cite{florian2021scalable}.
  \item Trajectory PMBM and PMB filters using clustering and assignment  \cite{xia2019extended}, referred to as TPMBM-CA and TPMB-CA.
  \item Trajectory PMBM and PMB filters using stochastic optimization \cite{soextended}, referred to as TPMBM-SO and TPMB-SO.
\end{itemize}

For all the TPMBM and TPMB filters, we consider the implementations for the set of all trajectories. We also present PMB-BP, which can be obtained from TPMB-BP by marginalizing out all the previous states after the prediction step \cite{granstrom2019poisson}, and its trajectory estimates are obtained by linking object state estimates that originate from the same first detection (i.e., the same Bernoulli component). Note that TPMB-BP and PMB-BP have the same filtering performance in terms of the set of current object states estimate. 

\subsection{Single object model}\label{sec_singleObjectModel}
There are several extended object models available in the literature, see \cite{extendedoverview} for an overview. We consider the random matrix model in \cite{randomMatrix2}, in which the object shape is approximated as an ellipse. The random matrix model has been used in several PMBM implementations \cite{pmbmextended2,xia2019extended,soextended,xia2021poisson}, thereby making the comparison easy.

The single-object state $x_k = (e_k,E_k)$, represented using a tuple \cite[Eq. (21.88)]{rfs}, consists of a kinematic state vector, which describes the two-dimensional position and velocity of the object, and an extent state $E_k$, which is a $2\times 2$ symmetric positive definite matrix. We assume that the object moves according to a nearly constant velocity model and its extent remains unchanged over time. In this case, the object state transition density is 
\begin{equation*}
  g_k(x_k|x_{k-1}) = {\cal N}\left(e_k;Fe_{k-1}+Q\right){\cal W}\left(E_k;E_{k-1}/q,q\right),
\end{equation*}
\begin{equation*}
  F = I_2 \otimes \begin{bmatrix}
    1 & T_s\\ 0 & 1
  \end{bmatrix}, \quad Q = \sigma_q^2I_2 \otimes \begin{bmatrix}
    T_s^3/3 & T_s^2/2 \\ T_s^2/2 & T_s
  \end{bmatrix}
\end{equation*}
where $I_2$ is a $2\times 2$ identity matrix, $\otimes$ denotes the Kronecker product, $T_s$ is the sampling period, $\sigma_q$ is standard acceleration deviation, and ${\cal W}\left(E_k;E_{k-1}/q,q\right)$ represents a Wishart distribution with mean $E_k$ and degree of freedom $q$. The single measurement likelihood is 
\begin{equation*}
  \ell_k(z_k|x_k) = {\cal N}(z_k;He_k,\rho E_k + R),
\end{equation*}
\begin{equation*}
  H = I_2 \otimes \begin{bmatrix}
    1 & 0
  \end{bmatrix}, \quad R = \sigma_r^2I_2
\end{equation*}
where $\rho>0$ is a scaling factor and $\sigma_r$ is the measurement noise deviation.

The gamma distribution is the conjugate prior for the Poisson likelihood, whereas the Gaussian inverse-Wishart (GIW) distribution is the conjugate prior for the multivariate Gaussian likelihood with unknown mean and covariance. Therefore, the single-object state density in TPMB(M)-CA and TPMB(M)-SO is a gamma GIW (GGIW).

\subsection{Simulation scenario}
We consider the same scenario as in \cite{florian2021scalable} where ten object tracks intersect at the centre of the region of interest of size $[-150~\text{m},150~\text{m}]\times[-150~\text{m},150~\text{m}]$. The ten objects start moving towards the centre from positions uniformly placed on a circle of radius $75~\text{m}$ around the centre with initial velocity $10~\text{m/s}$, and then they become closely-spaced for some time before they separate. The extent of each object is obtained by sampling from an inverse-Wishart distribution with mean $9I_2$ and degree of freedom $1000$. The object survival probability is $p^S = 0.99$. The following parameters for the single object model specified in Section \ref{sec_singleObjectModel} are used: $T_s=0.2$, $\sigma_q = 1$, $q = 1000$, $\rho = 1$ and $\sigma_r = 1$. The PPP clutter is uniformly distributed in the region of interest with mean $\gamma^C = 10$. Three different settings for the mean number of measurements in the Poisson spatial model are considered: $\gamma\in\{3,5,7\}$, and the corresponding effective probabilities of detection are approximately $1-\ell(\emptyset|\cdot)\in\{0.950,0.993,0.999\}$.

\begin{figure}[!t]
\centering
\includegraphics[width=0.8\linewidth]{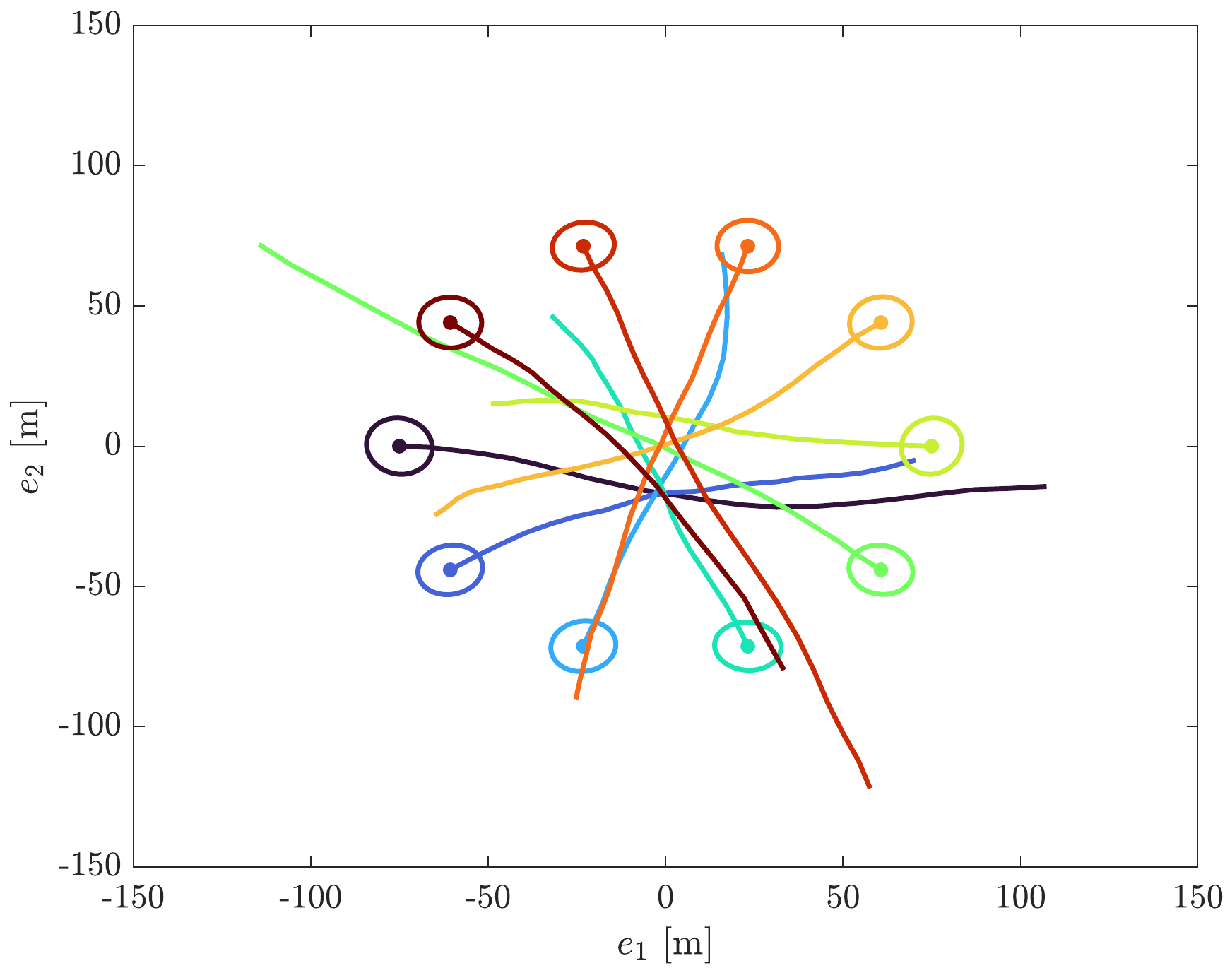}
\caption{Example realization of true object trajectories. True object extents at the times objects appear are also shown. The ten objects appear in pairs at time step $3,6,9,12,15$ and disappear also in pairs at time step $83,86,89,92,95$, respectively. The true object extents may (partly) overlap around the centre.}
\label{fig_groundTruth}
\end{figure}

\subsection{Implementation details}
For all the implementations, the Poisson rate of PPP birth is set to 0.01, and in the birth density, the velocity is Gaussian distributed with zero mean and covariance $100I_2$, and the extent is inverse-Wishart distributed with mean $9I_2$ and degree of freedom $1000$. For particle-based implementations, the position in the birth density is uniformly distributed in the region of interest, whereas for GGIW implementations, the position is Gaussian distributed with zero mean and covariance $150^2I_2$, and the Poisson measurement rate is gamma distributed with shape $1000\gamma$ and scale $1000$.

For implementations using particle BP, each particle describes both kinematic and extent states, and therefore more particles are needed to guarantee a reasonable performance as compared to particle-based implementation for point object tracking \cite{meyer2018message}. Moreover, new kinematic and extent states are obtained by sampling the Gaussian and Wishart distributions, respectively. We note that it is not difficult to generate random Wishart matrices. Empirical results show that the runtime taken by sampling from the Gaussian and Wishart distributions is rather marginal.

For PMB-BP and SPA, the number of particles is 2000, whereas for TPMB-BP the number of particles is adaptive, and we set $L_b = 2000$. Empirical results show that using 2000 particles is a good trade-off between runtime and estimation performance. For all these three implementations, the number of message passing iterations is set to 3, and the measurement-driven initialization of newly detected objects, as described in Section \ref{sec_measure_driven}, is used, where samples of state elements that do not depend on the measurements are drawn from the birth density. In addition, we use message censoring and measurement reordering, as discussed in \cite{florian2020scalable}, to facilitate track initialization and reduce computational complexity. Furthermore, we prune Bernoulli components with probability of existence smaller than $10^{-3}$. For TPMB-BP, we further discard particles using \eqref{eq_card_death} with threshold $10^{-4}$. 

For GGIW implementations, we prune global hypotheses (MBs) with weight smaller than $10^{-2}$ and Bernoulli components with probability of existence smaller than $10^{-3}$. We also prune GGIW components in the PPP intensity of undetected objects with weight smaller than $10^{-3}$. In addition, we use ellipsoidal gating with gate size $13.8$ to reduce computational complexity. For TPMB(M)-CA and TPMB(M)-SO, we consider $L$-scan implementation with $L=1$, i.e., no smoothing-while-filtering is performed. Furthermore, Bernoulli components with probability being alive at current time step smaller than $10^{-4}$ are not updated. For TPMB(M)-CA, we first apply the density-based spatial clustering of applications with noise (DB-SCAN) using 200 different distance values equally spaced between $0.1$ and $20$ to obtain a set of different measurement partitions, and then for each measurement partition and global hypothesis $a$, we apply Murty's algorithm\footnote{The C++ implementation in the Tracker Component Library \cite{crouse2017tracker} is used.} \cite{crouse2016implementing} to find the $\lceil 1000w_k^a \rceil$ best cluster-to-Bernoulli assignments. For TPMB(M)-SO, the number of iterations in SO at time step $k$ is set to $10m_k$.

For all the implementations, the object/trajectory estimates are extracted from Bernoulli components with probability of existence no smaller than $0.5$.

\subsection{Performance evaluation}\label{sec_performance_evaluation}
The trajectory estimation performance is evaluated using the linear programming (LP) metric $d(\cdot,\cdot)$ for sets of trajectories \cite{garcia2020metric} with parameters: cut-off distance $20$, order $1$, and track switch cost $2$, see Appendix \ref{appendix_e} for the detailed expression. The LP metric is integrated with the Gaussian Wasserstein distance (GWD) for performance evaluation of extended object estimates with ellipsoidal extent \cite{gwd}. In the simulated scenarios, we apply the metric at each time step, and normalize it by the time step. This enables a comparison of how the LP metric (filtering performance) evolves over time. 

We also apply the LP metric \cite{garcia2020metric} at the final time step to measure the smoothing error. Only the smoothing of kinematic states is performed, which is given by the backward simulation (with 100 iterations) and Rauch-Tung-Striebel smoother \cite{sarkka2013bayesian} for particle-based and GGIW implementations, respectively. 

\subsection{Results}

\begin{figure*}[!t]
  \centering
  \subfloat[$\gamma = 3$]{\includegraphics[width=0.33\linewidth]{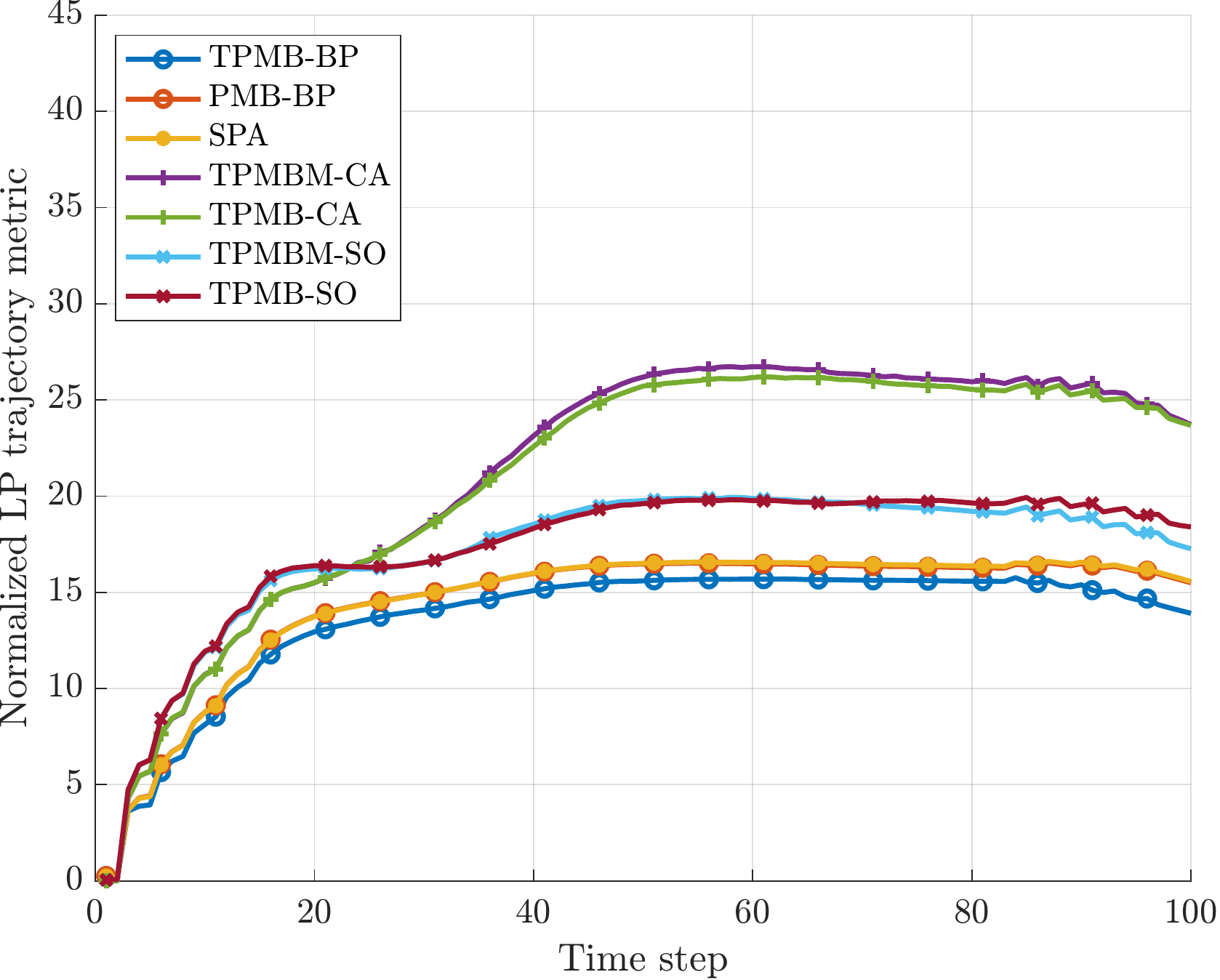}%
  \label{fig_trajFilterError3}}
  \subfloat[$\gamma = 5$]{\includegraphics[width=0.33\linewidth]{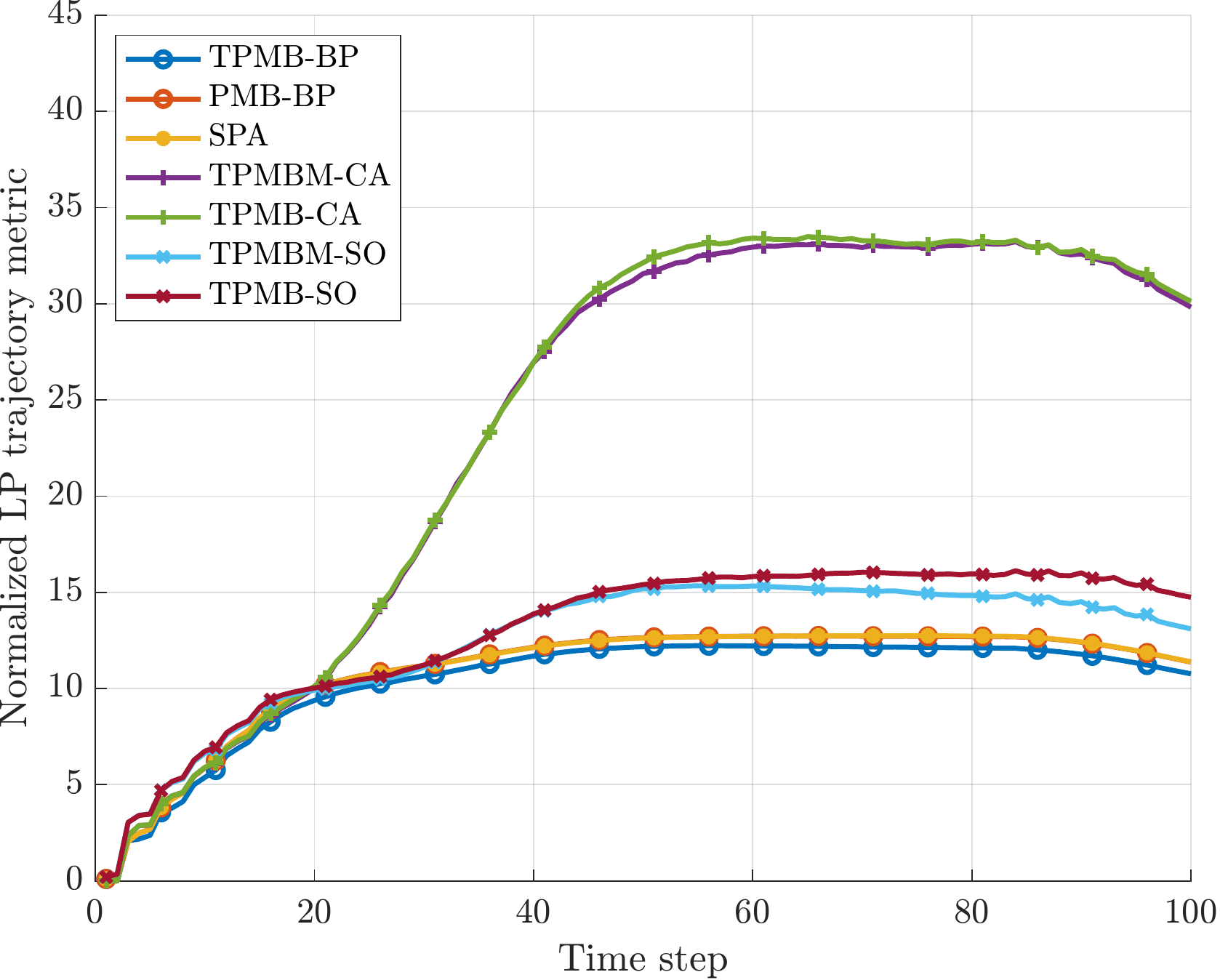}%
  \label{fig_trajFilterError5}}
  \subfloat[$\gamma = 7$]{\includegraphics[width=0.33\linewidth]{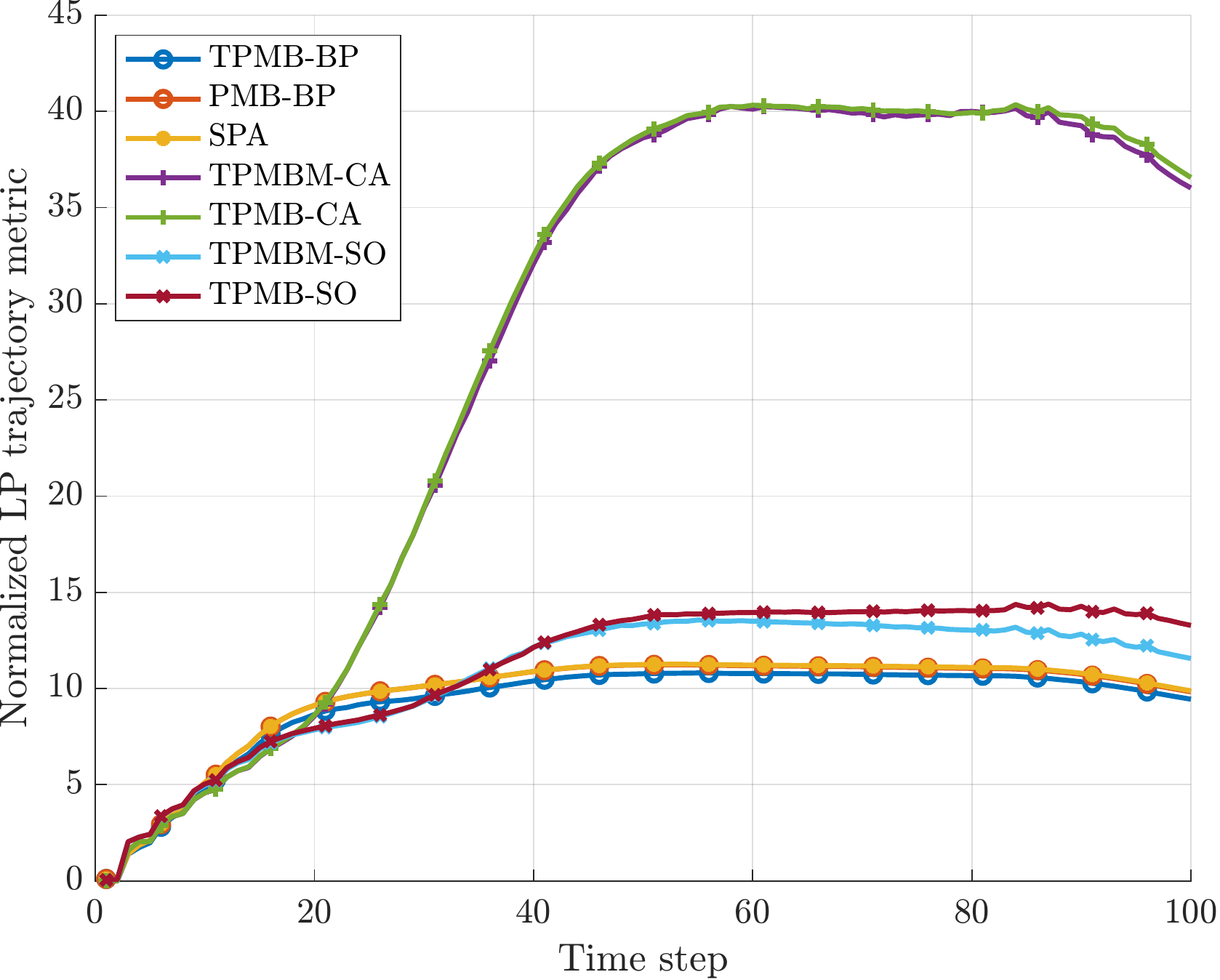}%
  \label{fig_trajFilterError7}}
  \caption{Filtering performance with $\gamma\in\{3,5,7\}$ in terms of the normalized LP trajectory metric over time. The line of PMB-BP almost overlaps the line of SPA.}
  \label{fig_trajFilterError}
\end{figure*}

\begin{figure*}[!t]
  \centering
  \subfloat[$\gamma = 3$]{\includegraphics[width=0.33\linewidth]{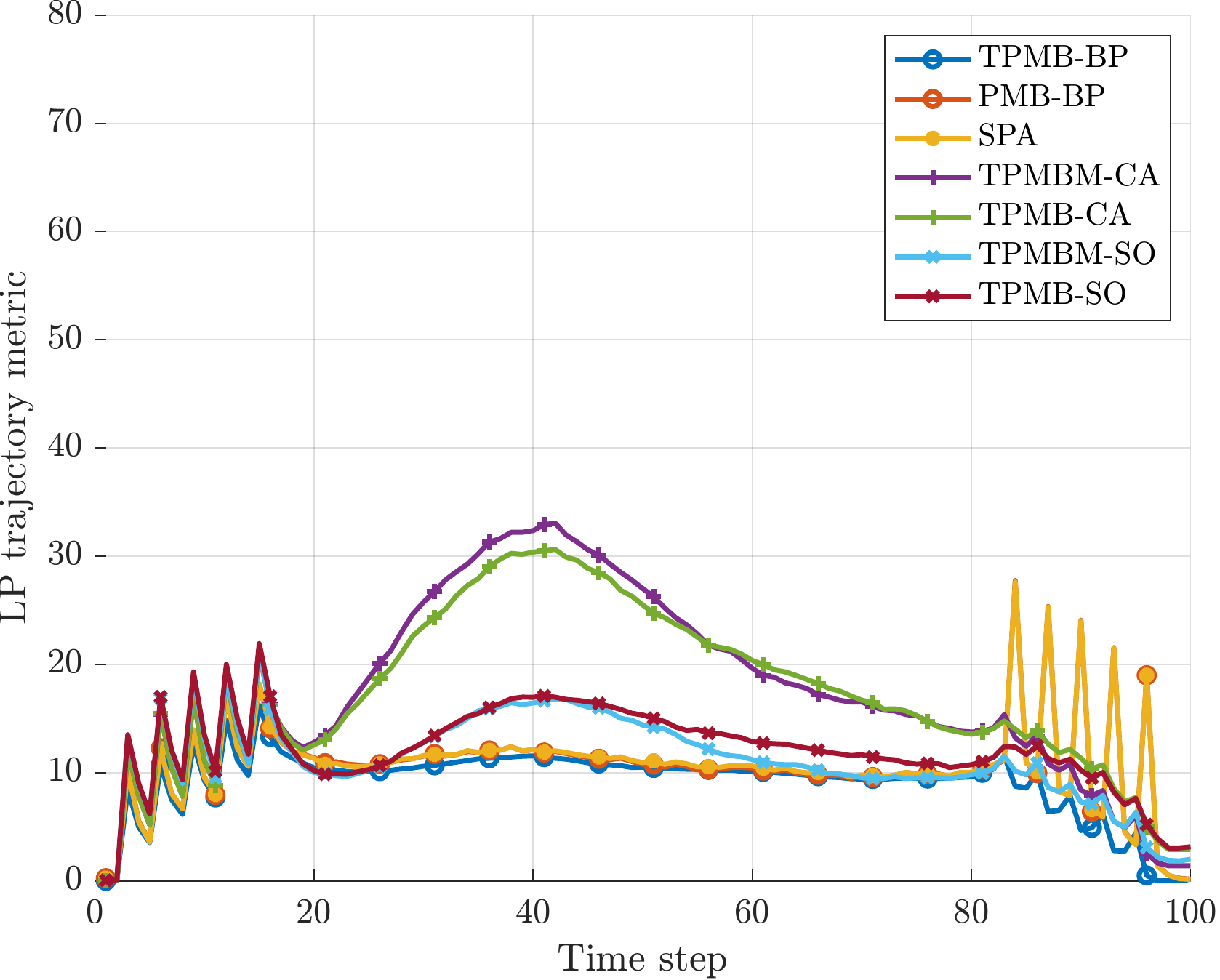}%
  \label{fig_trajSmootherError3}}
  \subfloat[$\gamma = 5$]{\includegraphics[width=0.33\linewidth]{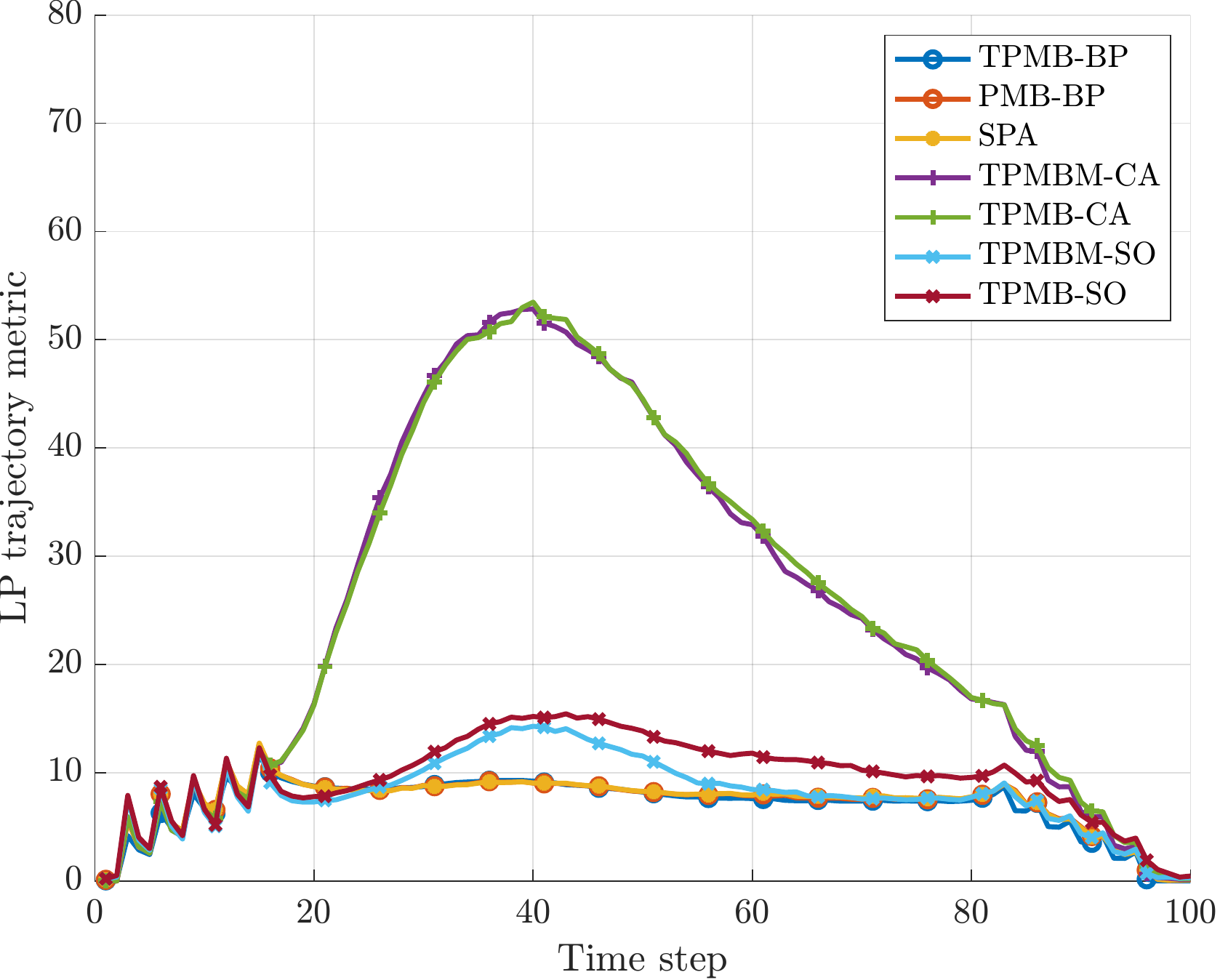}%
  \label{fig_trajSmootherError5}}
  \subfloat[$\gamma = 7$]{\includegraphics[width=0.33\linewidth]{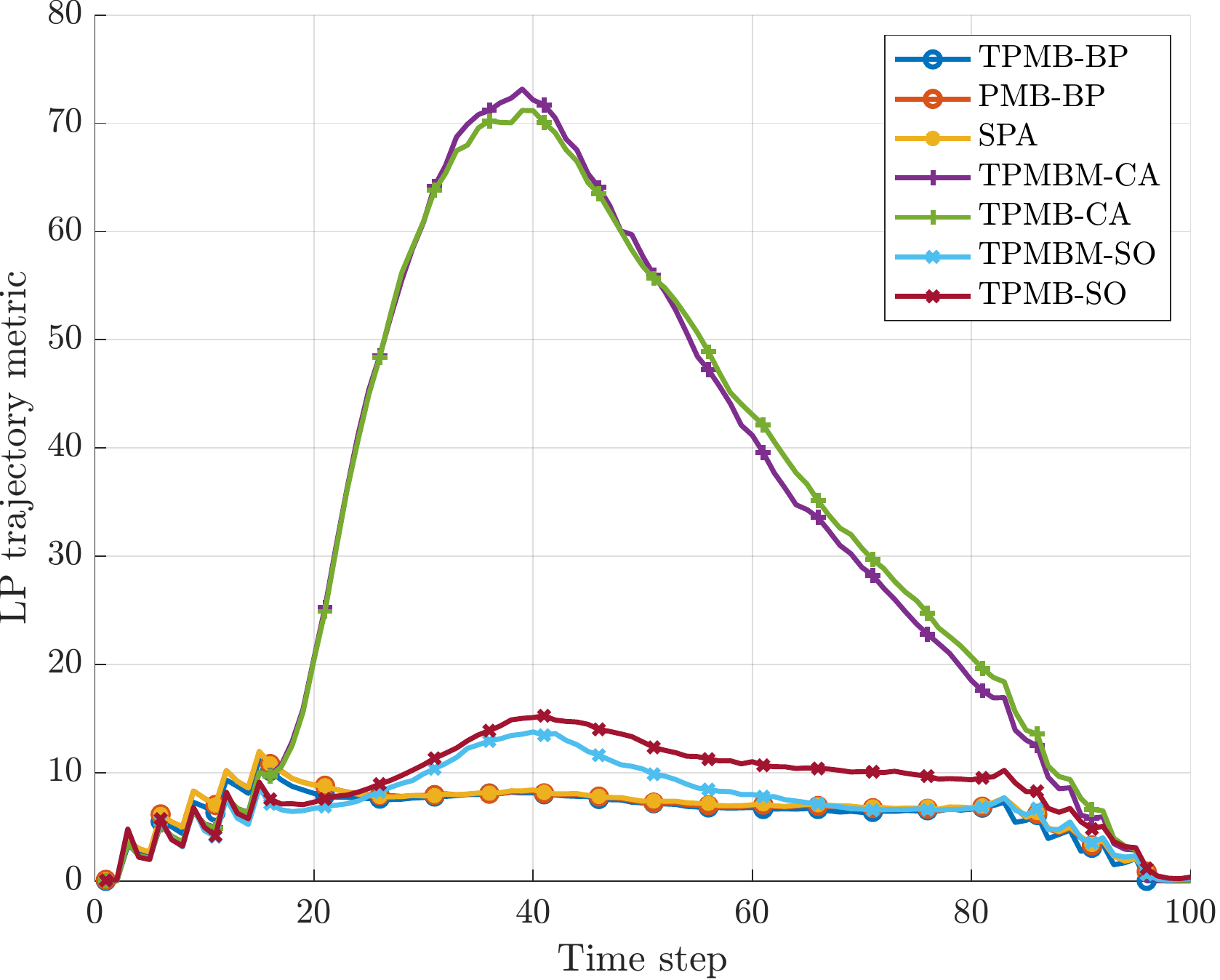}%
  \label{fig_trajSmootherError7}}
  \caption{Smoothing performance (evaluated at the final time step) with $\gamma\in\{3,5,7\}$ in terms of the LP trajectory metric over time. The line of PMB-BP almost overlaps the line of SPA.}
  \label{fig_trajSmootherError}
\end{figure*}

The filtering and smoothing estimation errors in terms of the LP trajectory metric are presented in Fig. \ref{fig_trajFilterError} and Fig. \ref{fig_trajSmootherError}, respectively. In addition, the decompositions of the trajectory metrics into costs due to state estimation error, missed and false detections, as well as track switches, are given in Table \ref{tab_table1} and Table \ref{tab_table2}. The results show that implementations using particle BP have the best estimation performance, followed by implementations using SO.

The different implementations have similar estimation performance when objects are well-spaced. When objects become closely-spaced, implementations using C\&A and SO  present increased estimation error, and in particular, implementations using C\&A suffer from false detection error. This is because the clustering algorithm does not yield reasonable results when objects are in proximity: the generated measurement partitions contain either many small clusters or a single big cluster. For TPMB(M)-CA in the simulated scenario, the hypotheses with many small clusters have higher likelihoods, and therefore they tend to overestimate the number of objects. This also explains why implementations using C\&A show increased estimation error when the Poisson measurement rate $\gamma$ increases, which is instead the opposite for implementations using SO or particle BP that avoid clustering. Compared to implementations using SO, implementations using particle BP enjoy a better trade-off between estimation performance and runtime. When the data association uncertainty is high, implementations using SO need a significant number of iterations in the sampling step to obtain hypotheses with high likelihoods. However, when the objects are relatively well-spaced, implementations using SO actually present better performance. This can be observed in Fig. \ref{fig_trajSmootherError} around time step $15$ to time step $20$, and the difference is most noticeable in the case $\gamma = 7$.

For the three different implementations using particle BP, PMB-BP slightly outperforms SPA, and the moderate performance advantage of PMB-BP over SPA can be explained by the fact that PMB-BP also propagates the density of undetected objects over time. TPMB-BP shows the best trajectory estimation performance since estimates of trajectory end time at earlier time steps may be improved at later time steps. This is most obvious for the case $\gamma = 3$ where both PMB-BP and SPA present increased false detection error when objects die. When $\gamma = 3$, the effective probability of detection $0.95$ is smaller than the object survival probability $0.99$. Therefore, it is likely that the object death events cannot be immediately reported by the multi-object filter. 

The average runtime\footnote{MATLAB implementations on a single core of an Intel(R) Xeon(R) Gold 6226R CPU @ 2.90GHz.} of different implementations is given in Table \ref{tab_table3}. As can be seen, PMB-BP and SPA are the fastest. TPMB-BP is slower than TPMB-SO, but it has significantly better estimation performance. It has been observed that a time-consuming part of the current implementation of TPMB-BP is the appending of particles representing dead trajectories in the prediction step via struct arrays. Possible improvements may be brought by using more efficient data structures.

\begin{table*}[!t]
    \caption{Filtering performance in terms of the normalized LP trajectory metric (summed over all the time steps) and its decomposition}
  \centering
  \resizebox{\textwidth}{!}{
    \begin{tabular}{c|ccccc|ccccc|ccccc}
    \hline
             & \multicolumn{5}{c|}{$\gamma=3$}          & \multicolumn{5}{c|}{$\gamma=5$}           & \multicolumn{5}{c}{$\gamma=7$}           \\ \hline
             & Total  & State   & Miss  & False & Switch & Total  & State   & Miss  & False  & Switch & Total  & State   & Miss & False  & Switch \\ \hline
    TPMB-BP  & $\underline{1365.1}$ & 1230.4 & 121.2 & 11.0  & 2.6    & $\underline{1039.0}$ & 1009.0 & 20.4  & 8.1    & 1.5    & $\underline{921.7}$  & 890.0  & 3.9  & 26.4   & 1.3    \\
    PMB-BP   & 1448.4 & 1222.2 & 175.0 & 48.2  & 3.1    & 1091.4 & 995.8  & 75.7  & 18.5   & 1.4    & 962.4  & 886.2  & 21.7 & 53.5   & 1.1    \\
    SPA      & 1452.1 & 1225.5 & 174.4 & 49.2  & 3.1    & 1091.4 & 996.4  & 75.4  & 18.1   & 1.4    & 964.1  & 886.1  & 23.2 & 54.4   & 1.1    \\
    TPMBM-CA & 2128.5 & 1394.2 & 337.4 & 379.7 & 17.2   & 2406.9 & 1258.0 & 99.4  & 1021.3 & 28.2   & 2827.4 & 1173.5 & 61.5 & 1553.9 & 38.5   \\
    TPMB-CA  & 2100.3 & 1374.9 & 319.0 & 390.1 & 16.3   & 2427.2 & 1236.8 & 86.2  & 1077.0 & 27.2   & 2846.9 & 1164.6 & 55.4 & 1589.3 & 37.6   \\
    TPMBM-SO & 1710.8 & 1247.8 & 410.6 & 45.8  & 6.6    & 1241.5 & 1042.9 & 145.8 & 48.0   & 4.9    & 1070.0 & 923.8  & 93.9 & 47.8   & 4.5    \\
    TPMB-SO  & 1726.1 & 1242.2 & 396.4 & 81.1  & 6.3    & 1296.6 & 1057.6 & 128.7 & 104.8  & 5.4    & 1120.7 & 941.8  & 81.8 & 92.3   & 4.9   
    \end{tabular}}
    \label{tab_table1}
\end{table*}

\begin{table*}[!t]
  \caption{Smoothing performance (evaluated at the final time step) in terms of the LP trajectory metric and its decomposition}
  \centering
  \resizebox{\textwidth}{!}{
  \begin{tabular}{c|ccccc|ccccc|ccccc}
  \hline    
           & \multicolumn{5}{c|}{$\gamma=3$}          & \multicolumn{5}{c|}{$\gamma=5$}           & \multicolumn{5}{c}{$\gamma=7$}         \\ \hline
           & Total   & State  & Miss  & False & Switch & Total  & State  & Miss & False  & Switch & Total   & State  & Miss & False  & Switch \\ \hline
  TPMB-BP  & $\underline{906.1}$  & 832.9 & 65.8  & 6.4   & 1.0     & $\underline{698.9}$  & 681.2 & 11.4 & 5.7    & 0.6    & $\underline{623.7}$   & 606.6 & 1.8  & 14.8   & 0.5    \\
  PMB-BP   & 1052.9 & 840.2 & 75.1  & 136.4 & 1.2     & 724.2  & 679.7 & 21.0 & 22.9   & 0.6    & 654.6   & 606.5 & 4.8  & 42.9   & 0.5    \\
  SPA      & 1058.0 & 843.3 & 76.3  & 137.2 & 1.1     & 725.7  & 683.0 & 19.6 & 22.4   & 0.6    & 659.8   & 606.7 & 7.7  & 44.9   & 0.5    \\
  TPMBM-CA & 1744.0 & 956.0 & 177.8 & 596.8 & 13.4    & 2464.0 & 900.2 & 56.5 & 1484.7 & 22.5   & 3140.8  & 867.3 & 39.9 & 2201.8 & 31.8   \\
  TPMB-CA  & 1726.7 & 911.5 & 161.3 & 643.2 & 10.6    & 2483.5 & 859.2 & 47.1 & 1557.5 & 19.5   & 3184.4  & 845.3 & 35.4 & 2275.7 & 28.0   \\
  TPMBM-SO & 1117.5 & 820.7 & 212.3 & 79.7  & 4.9     & 810.2  & 671.7 & 83.7 & 51.0   & 3.7    & 723.0   & 602.5 & 60.4 & 56.8   & 3.4    \\
  TPMB-SO  & 1215.6 & 789.2 & 194.2 & 228.7 & 3.5     & 959.9  & 672.0 & 66.2 & 218.4  & 3.2    & 886.5   & 618.3 & 46.5 & 217.8  & 3.9   
  \end{tabular}}
  \label{tab_table2}
\end{table*}

\begin{table}[!t]
  \caption{Average runtime (in seconds) for a complete Monte Carlo run of different implementations}
  \centering
  \begin{tabular}{c|cllllcllllcllll}
  \hline
           & \multicolumn{5}{c}{$\gamma=3$} & \multicolumn{5}{c}{$\gamma=5$} & \multicolumn{5}{c}{$\gamma=7$} \\ \hline
  TPMB-BP  & \multicolumn{5}{c}{88.2}       & \multicolumn{5}{c}{86.6}       & \multicolumn{5}{c}{107.5}       \\
  PMB-BP   & \multicolumn{5}{c}{26.5}       & \multicolumn{5}{c}{42.2}       & \multicolumn{5}{c}{60.6}       \\
  SPA      & \multicolumn{5}{c}{26.1}       & \multicolumn{5}{c}{42.2}       & \multicolumn{5}{c}{60.7}       \\
  TPMBM-CA & \multicolumn{5}{c}{144.7}      & \multicolumn{5}{c}{304.4}      & \multicolumn{5}{c}{452.8}      \\
  TPMB-CA  & \multicolumn{5}{c}{103.4}      & \multicolumn{5}{c}{236.3}      & \multicolumn{5}{c}{375.4}      \\
  TPMBM-SO & \multicolumn{5}{c}{699.3}      & \multicolumn{5}{c}{1330.9}     & \multicolumn{5}{c}{2081.0}     \\
  TPMB-SO  & \multicolumn{5}{c}{34.7}       & \multicolumn{5}{c}{60.5}       & \multicolumn{5}{c}{91.6}      
  \end{tabular}
  \label{tab_table3}
  \vspace{-5mm}
\end{table}

\section{Conclusion}
In this paper, we present a PMBM conjugate prior on the posterior of sets of trajectories for a generalized measurement model. We also present the factor graph representation of the joint posterior of the PMBM set of trajectories and association variables for the Poisson spatial measurement model. Based on these important theoretical contributions, we present two TPMB filters for multiple EOT implemented using particle BP: one estimates the set of alive trajectories, and the other estimates the set of all trajectories. The proposed implementations show excellent performance advantages when objects move in proximity, while also providing full trajectory information. 

For the future work, it would be interesting to investigate how to extend the current work to consider multi-scan data associations \cite{williams2018multiple} and tracking co-existing point and extended objects \cite{garcia2021poisson}. Another interesting future work direction is to study how to integrate the particle flow filter \cite{mori2022note}, which does not suffer from the particle path degeneracy problem, into filters based on sets of trajectories, such that one can directly extract trajectory estimates from the multi-trajectory posterior density.

\bibliographystyle{IEEEtran}
\bibliography{mybibli.bib}



\cleardoublepage
{\bfseries \huge Supplementary Materials}

{\appendices
\section{}\label{appendix_a}
In this appendix, we present a short proof of Proposition \ref{prop_update}.

It has been shown in \cite[Theorem 1]{garcia2021poisson} that the update of a PMBM prior with single-object state in a LCHS space and the generalized measurement model in Section \ref{sec_clutter} is also a PMBM. Furthermore, given that the single-object state space is LCHS, the single-trajectory space $T_{(k^\prime)}$ is also LCHS \cite[Appendix A]{garcia2019multiple}. This means that finite set statistics \cite{rfs} can be used on sets of trajectories. Therefore, Proposition \ref{prop_update} can be understood as an extension of the PMBM update for sets of objects  \cite[Theorem 1]{garcia2021poisson} to sets of trajectories.

\section{}\label{appendix_factor_graph}
In this appendix, we prove Proposition \ref{proposition_factor_graph}. In what follows, we first present the joint posterior of set of trajectories and global hypothesis augmented with auxiliary variables, which does not involve the summation over set partitions (cf. \eqref{eq_pmbm_au}). Then, we proceed to describe how to rewrite the joint posterior using the object-oriented and measurement-oriented association variables \cite{florian2020scalable}. At last, we show how to further simplify such joint posterior to exclude the object-oriented association variables.

\subsection{Joint posterior of trajectories and global hypothesis}
Given a predicted TPMB density at time step $k$ of the form \eqref{eq_pmb_predict} and measurements ${\bf z}_k$, the joint posterior of set of trajectory and global hypothesis is 
\begin{align}\label{eq_joint_posterior}
  \widetilde{f}_{k|k}(\widetilde{{\bf X}}_k,a) &= \widetilde{f}_{k|k}(\widetilde{{\bf X}}_k|a)w^a_{k|k}\psi(a)\nonumber\\
  &= w^a_{k|k} \widetilde{f}^p_{k|k}(\widetilde{{\bf Y}}_k)\prod_{i=1}^{n_{k|k}}\widetilde{f}^{i,a^i}_{k|k}\left(\widetilde{{\bf X}}_k^i\right)\psi(a)\nonumber\\
  &\propto \widetilde{f}^p_{k|k}(\widetilde{{\bf Y}}_k)\prod_{i=1}^{n_{k|k}}w^{i,a^i}_{k|k}\widetilde{f}^{i,a^i}_{k|k}\left(\widetilde{{\bf X}}^i_k\right)\psi(a)\nonumber\\
  &=  \widetilde{f}^p_{k|k}(\widetilde{{\bf Y}}_k) \prod_{i=1}^{n_{k|k}}g^{i,a^i}_{k|k}\left(\widetilde{{\bf X}}^i_k\right)\psi(a)
\end{align}
where the constraint on global hypothesis $a$ can be expressed as an indicator function 
\begin{equation}
  \psi(a) = \begin{cases}
    1 & \biguplus_{i=1}^{n_{k|k}}{\cal M}_{k}^{i,a^i} = {\cal M}_{k}\\
    0 & \text{otherwise},
  \end{cases}\label{eq_constraint}
\end{equation}
and we introduce the unnormalized Bernoulli densities $g^{i,a^i}_{k|k}(\cdot)$, each of which is given by the product of a local hypothesis density $\widetilde{f}^{i,a^i}_{k|k}(\widetilde{{\bf X}}^i_k)$ and its corresponding weight $w^{i,a^i}_{k|k}$. Using their expressions presented in Proposition \ref{prop_update}, we have, for $i \in \{1,\dots,n_{k|k-1}\}$, 
\begin{align}
  &g^{i,1}_{k|k}\left(\widetilde{{\bf X}}\right)\nonumber\\ &= \begin{cases}
    r^i_{k|k-1}\ell_k(\emptyset|X)f^i_{k|k-1}(X)\delta_i[u] & \widetilde{{\bf X}} = \{(u,X)\}\\
    1 - r^i_{k|k-1} & \widetilde{{\bf X}} = \emptyset\\
    0 & \text{otherwise},
  \end{cases}\label{eq_unnormalizeber1}
\end{align}
for $i \in \{1,\dots,n_{k|k-1}\}$, $j\in\{2,\dots,2^{m_k}-1\}$, $a^i = j+1$, 
\begin{align}
  &g_{k|k}^{i,a^i}\left(\widetilde{{\bf X}}\right)\nonumber\\ &= \begin{cases}
    r_{k|k-1}^i\ell_k\left({\bf w}_k^{j}|X\right)f^i_{k|k-1}(X)\delta_i[u] & \widetilde{{\bf X}} = \{(u,X)\}\\
    0 & \text{otherwise},
  \end{cases}\label{eq_unnormalizeber2}
\end{align}
for $i\in\{n_{k|k-1}+1,\dots,n_{k|k-1}+m_k\}$,
\begin{equation}
  \label{eq_nonexistnew}
  g^{i,1}_{k|k}\left(\widetilde{{\bf X}}\right) = \begin{cases}
    1 & \widetilde{{\bf X}} = \emptyset\\
    0 & \text{otherwise},
  \end{cases}
\end{equation}
and for $i = n_{k|k-1}+j,j\in\{1,\dots,m_k\},\iota\in \{1,\dots,2^{j-1}\}$, $a^i = \iota+1$
\begin{align}
  &g^{i,a^i}_{k|k}\left(\widetilde{{\bf X}}\right)\nonumber\\ &= \begin{cases}
    \ell_k\left({\bf w}_k^{i,\iota}|X\right)\lambda_{k|k-1}(X)\delta_i[u] & \widetilde{{\bf X}} = \{(u,X)\}\\
    \lambda_k^C\left(z_k^j\right) & \widetilde{{\bf X}} = \emptyset,{\bf w}_k^{i,\iota} = \left\{z_k^j\right\}\\
    0 & \text{otherwise}.
  \end{cases}\label{eq_unnormalizeber3}
\end{align}

\subsection{Data association variables}
Until now, we have used variable $a\in {\cal A}_{k|k}$ to represent the global hypotheses. To obtain a neat factor graph representation of the joint posterior, it is useful to describe the Bernoulli-to-measurement association ${\cal M}_k^{i,a^i}$ using association vectors. 
We first introduce the binary object-oriented association vector $\alpha_k = [{\alpha^1_k}^T,\dots,{\alpha_k^{n_{k|k}}}^T]^T$ where 
\begin{equation*}
  \alpha_k^i = \begin{cases}
    \left[\alpha_k^{i,1},\dots,\alpha_k^{i,m_k}\right]^T, & i\in\{1,\dots,n_{k|k-1}\}\\
    \left[\alpha_k^{i,1},\dots,\alpha_k^{i,j}\right]^T, & i = n_{k|k-1} + j, j\in\{1,\dots,m_k\}
  \end{cases}
\end{equation*}
and $\alpha_k^{i,j}=1$ if and only if the $j$-th measurement $z_k^j$ is associated to the $i$-th potential object (described by the $i$-th Bernoulli component). Note that the object-oriented association vector corresponding to the $j$-th new Bernoulli component has length $j$. This is a direct result of the local hypothesis representation for new Bernoulli components specified in Proposition \ref{prop_update}.

For TPMB filters\footnote{For the TPMB filter, the global hypothesis does not include the measurement association history.}, from global hypothesis $a\in{\cal A}_{k|k}$ \eqref{eq_globalhypo} we can represent the data associations at time step $k$ by object-oriented or measurement-oriented association variables; the latter has been introduced in Section \ref{sec_factor_graph_formulation}. Using a hybrid representation for both object-oriented and measurement-oriented association variables makes it possible for developing many scalable MOT algorithms using BP \cite{meyer2018message}. Later in Section \ref{sec_simplified_fg}, we will show that, for EOT with Poisson spatial model, the factor graph constructed using the hybrid association vectors can be simplified by marginalizing out the object-oriented association vector.

\subsection{Joint posterior of trajectories and association variables}

We start by rewriting the unnormalized Bernoulli densities $g^{i,a^i}_{k|k}(\cdot)$ as functions of the object-oriented association vectors $\alpha_k^i$. This is done by plugging \eqref{eq_measliktra} into \eqref{eq_unnormalizeber1}, \eqref{eq_unnormalizeber2} and \eqref{eq_unnormalizeber3}, which yields, for $i\in\{1,\dots,n_{k|k-1}\}$,
\begin{equation}\label{eq_g_n}
  g^{i,a^i}_{k|k}\left(\widetilde{{\bf X}}\right) = \underline{f}^{i}_{k|k-1}\left(\widetilde{{\bf X}}\right)\prod_{j=1}^{m_k}\hat{q}_k\left(\widetilde{{\bf X}},\alpha_k^{i,j};z_k^j\right),
\end{equation}
and for $i = n_{k|k-1}+j$, $j\in\{1,\dots,m_k\}$,
\begin{equation}\label{eq_g_m}
  g^{i,a^i}_{k|k}\left(\widetilde{{\bf X}}\right) = \overline{f}^i_{k|k-1}\left(\widetilde{{\bf X}}\right)\hat{v}_k\left(\widetilde{{\bf X}},\alpha_k^{i,j};z_k^j\right)\prod_{l=1}^{j-1}\hat{q}_k\left(\widetilde{{\bf X}},\alpha_k^{i,j};z_k^j\right)
\end{equation}
where the product over function $\hat{q}_k$ in \eqref{eq_g_m} reduces to $1$ when $j=1$, $\underline{f}^{i}_{k|k-1}(\cdot)$ and $\overline{f}^i_{k|k-1}(\cdot)$ are respectively given in \eqref{eq_existingprior} and \eqref{eq_newbornprior}, and 
\begin{align}
  &\hat{q}_k\left(\widetilde{{\bf X}},\alpha_k^{i,j};z_k^j\right)\nonumber\\ &= \begin{cases}
    \ell_k\left(z_k^j|X\right)\gamma_k(X)\delta_i[u] & \widetilde{{\bf X}} = \{(u,X)\},\alpha_k^{i,j}=1\\
    1 & \alpha_k^{i,j} = 0\\
    0 & \text{otherwise},
  \end{cases}\\
  &\hat{v}_k\left(\widetilde{{\bf X}},\alpha_k^{i,j};z_k^j\right)\nonumber\\ &= \begin{cases}
    \ell_k\left(z_k^j|X\right)\gamma_k(X)\delta_i[u] & \widetilde{{\bf X}} = \{(u,X)\},\alpha_k^{i,j} = 1\\
    \lambda_k^C\left(z_k^j\right) & \widetilde{{\bf X}} = \emptyset,\alpha_k^{i,j} = 0\\
    1 & \widetilde{{\bf X}} = \emptyset,\alpha_k^{i,j} = 0, \text{and} \\
      & \exists!i\in\{1,\dots,n_{k|k-1}\}: a_k^{i,j} = 1 \\
    0 & \text{otherwise}.
  \end{cases}\label{eq_absorb}
\end{align}
The third term in \eqref{eq_absorb} is a result of \eqref{eq_nonexistnew}, which represents the case that the $j$-th measurement is associated to an existing potential object and the corresponding new Bernoulli component has zero existence probability.

We further observe that conditioned on the measurements ${\bf z}_k$, the product $\prod_{z_k\in{\bf z}_k}\lambda^C_k(z_k)$ is a constant. If we divide \eqref{eq_joint_posterior} by this constant, the term $\lambda_k^C(\cdot)$ in \eqref{eq_absorb} reduces to $1$ and can be combined with the third term in \eqref{eq_absorb} without changing the proportionality of \eqref{eq_joint_posterior}, which yields
\begin{align}
  &q_k\left(\widetilde{{\bf X}},\alpha_k^{i,j};z_k^j\right)\nonumber\\ &= \begin{cases}
    \frac{\ell_k\left(z_k^j|X\right)\gamma_k(X)}{\lambda_k^C\left(z_k^j\right)}\delta_i[u] & \widetilde{{\bf X}} = \{(u,X)\},\alpha_k^{i,j}=1\\
    1 & \alpha_k^{i,j} = 0\\
    0 & \text{otherwise},
  \end{cases}\label{eq_q}\\
  &v_k\left(\widetilde{{\bf X}},\alpha_k^{i,j};z_k^j\right)\nonumber\\ &= \begin{cases}
    \frac{\ell_k\left(z_k^j|X\right)\gamma_k(X)}{\lambda_k^C\left(z_k^j\right)}\delta_i[u] & \widetilde{{\bf X}} = \{(u,X)\},\alpha_k^{i,j} = 1\\
    1 & \widetilde{{\bf X}} = \emptyset, \alpha_k^{i,j} = 0\\
    0 & \text{otherwise}.
  \end{cases}\label{eq_v}
\end{align}

Using the above results, we can rewrite the joint posterior of trajectories and global hypothesis \eqref{eq_joint_posterior} as the joint posterior of trajectories and object-oriented association vector 
\begin{align}\label{eq_joint_posterior2}
  &\widetilde{f}_{k|k}\left(\widetilde{{\bf X}}_k,\alpha_k\right)\nonumber\\ &\propto \widetilde{f}^p_{k|k}\left(\widetilde{{\bf Y}}_k\right)\prod_{i=1}^{n_{k|k-1}}\Biggl[\underline{f}^{i}_{k|k-1}\left(\widetilde{{\bf X}}_k^i\right)\prod_{j=1}^{m_k}q_k\left(\widetilde{{\bf X}}_k^i,\alpha_k^{i,j};z_k^j\right)\Biggr]\nonumber\\ &~~\times \prod_{i=n_{k|k-1}+1}^{n_{k|k}} \Biggl[\overline{f}^i_{k|k-1}\left(\widetilde{{\bf X}}^i_k\right)v_k\left(\widetilde{{\bf X}}_k^i,\alpha_k^{i,i-n_{k|k-1}};z_k^{i-n_{k|k-1}}\right)\nonumber\\
  &~~\times\prod_{j=1}^{i-n_{k|k-1}-1}q_k\left(\widetilde{{\bf X}}_k^i,\alpha_k^{i,j};z_k^j\right)\Biggr]\zeta(\alpha_k)
\end{align}
where $\zeta(\alpha_k)$ is an indicator function (cf. \eqref{eq_constraint}) used to express the constraint on $\alpha_k$. We note that $\zeta(\alpha_k)$ is a function of all the binary variables $\alpha_k^{i,j}$, and thus it is not efficient to run BP directly on the factor graph constructed using \eqref{eq_joint_posterior2}. To solve this problem, a computationally feasible solution is to stretch $\zeta(\alpha_k)$ using the following binary indicator function, which checks consistency for any pair $(\alpha_k^{i,j},\beta_k^j)$ of object-oriented and measurement-oriented association variables \cite{florian2020scalable}:
\begin{equation}\label{eq_consistent}
  \Psi\left(\alpha_k^{i,j},\beta_k^j\right) = \begin{cases}
    0 & \alpha_k^{i,j} = 1,\beta_k^j\neq i\text{~or~}\alpha_k^{i,j} = 0,\beta_k^j = i\\
    1 & \text{otherwise}
  \end{cases}
\end{equation}
for $i\in\{1,\dots,n_{k|k}\}$, $j\in\{1,\dots,m_k\}$. Specifically, this yields the following expression of the joint posterior of trajectories and association variables 
\begin{align}\label{eq_factor_graph1}
  &\widetilde{f}_{k|k}\left(\widetilde{{\bf X}}_k,\alpha_k,\beta_k\right)\nonumber\\ &\propto \widetilde{f}^p_{k|k}\left(\widetilde{{\bf Y}}_k\right)\prod_{i=1}^{n_{k|k-1}}\Biggl[\underline{f}^{i}_{k|k-1}\left(\widetilde{{\bf X}}_k^i\right)\prod_{j=1}^{m_k}q_k\left(\widetilde{{\bf X}}_k^i,\alpha_k^{i,j};z_k^j\right)\nonumber\\ &~~\times\Psi\left(\alpha_k^{i,j},\beta_k^j\right)\Biggr] \prod_{i=n_{k|k-1}+1}^{n_{k|k}} \Biggl[\overline{f}^i_{k|k-1}\left(\widetilde{{\bf X}}^i_k\right)\nonumber\\ &~~\times v_k\left(\widetilde{{\bf X}}_k^i,\alpha_k^{i,i-n_{k|k-1}};z_k^{i-n_{k|k-1}}\right)\nonumber\\
  &~~\times\Psi\left(\alpha_k^{i,i-n_{k|k-1}},\beta_k^{i-n_{k|k-1}}\right)\nonumber\\
  &~~\times\prod_{j=1}^{i-n_{k|k-1}-1}q_k\left(\widetilde{{\bf X}}_k^i,\alpha_k^{i,j};z_k^j\right)\Psi\left(\alpha_k^{i,j},\beta_k^j\right)\Biggr],
\end{align}
and its corresponding factor graph representation is illustrated in Fig. \ref{fig_factor_graph1}. 

\begin{figure}[!t]
\centering
\includegraphics[width=\columnwidth]{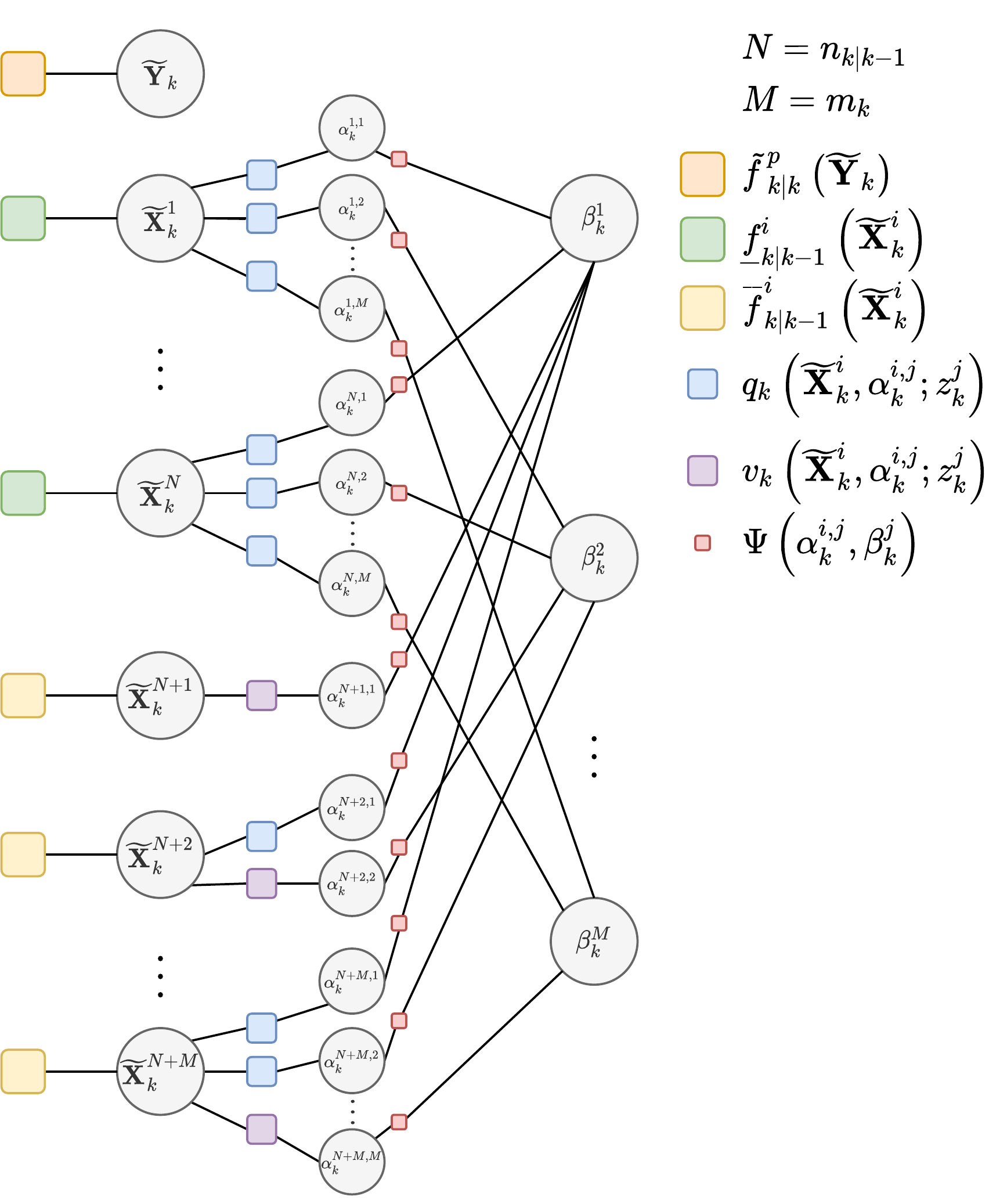}
\caption{Factor graph for the factorization in \eqref{eq_factor_graph1} where variable nodes are represented using square and factor nodes are represented using circle.}
\label{fig_factor_graph1}
\end{figure}

\subsection{A simplified representation of the joint posterior}\label{sec_simplified_fg}

For the factor graph shown in Fig. \ref{fig_factor_graph1}, each object-oriented association variable node $\alpha_k^{i,j}$ is connected to two factor nodes, one connecting $\alpha_k^{i,j}$ to a Bernoulli variable node $\widetilde{{\bf X}}^i$ and the other connecting $\alpha_k^{i,j}$ to a measurement-oriented association variable node $\beta_k^j$. This means that each variable node $\alpha_k^{i,j}$ and its two neighbouring factor nodes can be combined into a single factor node that directly connects variable nodes $\widetilde{{\bf X}}^i$ and $\beta_k^j$, by marginalizing out $\alpha_k^{i,j}$ from the factorization \eqref{eq_factor_graph1}. In particular, we have 
\begin{equation*}
  \widetilde{f}_{k|k}\left(\widetilde{{\bf X}}_k,\beta_k\right) = \sum_{\alpha_k^{i,j}\in\{0,1\}}\widetilde{f}_{k|k}\left(\widetilde{{\bf X}}_k,\alpha_k,\beta_k\right),
\end{equation*}
whose final expression is given by \eqref{eq_factor_graph2}, and \eqref{eq_s_underline} and \eqref{eq_s_overline} can be obtained by 
\begin{align}
  \underline{s}_k\left(\widetilde{{\bf X}}_k^i,\beta_k^j;z_k^j\right) &= \sum_{\alpha_k^{i,j}\in\{0,1\}}q_k\left(\widetilde{{\bf X}}_k^i,\alpha_k^{i,j};z_k^j\right)\Psi\left(\alpha_k^{i,j},\beta_k^j\right),\nonumber\\
  \overline{s}_k\left(\widetilde{{\bf X}}_k^i,\beta_k^j;z_k^j\right) &= \sum_{\alpha_k^{i,j}\in\{0,1\}}v_k\left(\widetilde{{\bf X}}_k^i,\alpha_k^{i,j};z_k^j\right)\Psi\left(\alpha_k^{i,j},\beta_k^j\right).\nonumber
\end{align}

This finishes the proof of Proposition \ref{proposition_factor_graph}.

\section{}\label{appendix_c}
In this appendix, we give the expressions of the predicted number of undetected objects at time step $k$ for both Bayesian solution and the solution in \cite{florian2021scalable} under the assumption that both the Poisson measurement rate $\gamma$ and survival probability $p^S$ are state-independent, and that both the Poisson measurement rate $\gamma$ and the Poisson birth rate $\overline{\lambda}^B$ are time-invariant.

Given the multi-object models in Section \ref{sec_model}, the Bayesian solution is given by \cite{rfs}
\begin{align}
  \overline{\lambda}^u_{k|k-1} &= \overline{\lambda}^B\left(1+p^Se^{-\gamma}+\cdots+\left(p^Se^{-\gamma}\right)^{k-1}\right)\nonumber\\
  &= \overline{\lambda}^B\frac{1-\left(p^Se^{-\gamma}\right)^k}{1-p^Se^{-\gamma}}.\label{eq_appendixb1}
\end{align}
The solution in \cite{florian2021scalable} is based on the assumption that newborn objects are always effectively detected (with probability one). This means that the probability that newborn objects generate zero measurement is zero. Thus, the probability mass function of Poisson measurement rate needs to be truncated to exclude the zero-measurement case and normalized by the probability $(1-e^{-\gamma})$ of generating non-zero measurement. In this case, the predicted number of undetected objects at time step $k$ is a constant \cite{florian2021scalable}
\begin{equation}
  \overline{\lambda}^u_{k|k-1} = \overline{\lambda}^B\frac{1}{1-e^{-\gamma}}.\label{eq_appendixb2}
\end{equation}
It can be verified that \eqref{eq_appendixb2} is an upper bound of \eqref{eq_appendixb1}, and that the bound becomes tighter as $k$ increases. A lower bound of the bias of \eqref{eq_appendixb2} is given by 
\begin{equation*}
  \frac{\overline{\lambda}^B\left(1-p^S\right)e^{-\gamma}}{\left(1-e^{-\gamma}\right)\left(1-p^Se^{-\gamma}\right)}.
\end{equation*}

\section{}\label{appendix_d}
The backward simulation particle smoother works by simulating individual trajectories backward in time, starting from the last time step. Many variants of the backward simulation particle smoother exist \cite{lindsten2013backward}. In the simulation, we use the one described in \cite[Algorithm 4]{lindsten2013backward}. Here, we describe how to apply backward simulation to the sets of particles representing the marginal single-object states in the time interval of interest, as described in Section \ref{sec_estimation}. The pseudocode in Algorithm \ref{alg:alg1} can be applied to the filtering densities of PMB-BP and SPA with minor modification. Specifically, when there are gaps in the trajectory, the missing state estimate is given by applying the inverse dynamic model $g^{-1}(\cdot|\cdot)$ to valid state estimate at future time steps.

\begin{algorithm}[!h]
  \caption{Backward simulation particle smoother.}\label{alg:alg1}
  \begin{algorithmic}[1]
  \REQUIRE $\left\{\left(w_{k^\prime|k^\prime}^{i,(l)},\overline{\chi}_{k^\prime|k^\prime}^{i,(l)}\right): l \in \mathbb{L}^i_{k^\prime} \right\}$ for $k^\prime = \hat{t}^i_s,\dots,\hat{t}^i_e$, $P$ 
  \ENSURE Trajectory estimate $\widehat{\chi}^i = \left( \hat{x}_{\hat{t}^i_s}^i,\dots,\hat{x}_{\hat{t}^i_e}^i \right)$.
  \STATE Normalize $\left\{w_{\hat{t}^i_e|\hat{t}^i_e}^{i,(l)}\right\}_{l\in\mathbb{L}_{\hat{t}^i_e}^i}$ to obtain $\left\{\hat{w}_{\hat{t}^i_e|\hat{t}^i_e}^{i,(l)}\right\}_{l\in\mathbb{L}_{\hat{t}^i_e}^i}$.
  \STATE Sample $\left\{ b_{\hat{t}^i_e}(j) \right\}_{j=1}^P$ from $\text{Categorical}\left(\left\{\hat{w}_{\hat{t}^i_e|\hat{t}^i_e}^{i,(l)}\right\}_{l\in\mathbb{L}_{\hat{t}^i_e}^i}\right)$.
  \STATE Set $\widetilde{x}_{\hat{t}^i_e}^j$ to $\overline{\chi}_{k^\prime|k^\prime}^{i,(b_{\hat{t}^i_e}(j))}$ for $j=1,\dots,P$.
  \STATE $\hat{x}_{\hat{t}^i_e}^i = \frac{1}{P}\sum_{j=1}^P \widetilde{x}_{\hat{t}^i_e}^j$.
  \FOR {$k^\prime = \hat{t}^i_e-1$ to $\hat{t}^i_s$}
  \FOR {$j=1$ to $P$}
  \STATE Compute $\hat{w}^{l,j} \propto g_{k^\prime+1}\left(\widetilde{x}_{k^\prime+1}^j|\overline{\chi}_{k^\prime|k^\prime}^{i,(l)}\right)~\forall~l\in\mathbb{L}_{k^\prime}^i$ and normalize.
  \STATE Draw $b_{k^\prime}(j)\sim\text{Categorical}\left(\left\{ \hat{w}^{l,j} \right\}_{l\in\mathbb{L}_{k^\prime}^i}\right)$.
  \STATE Set $\widetilde{x}^j_{k^\prime}$ to $\overline{\chi}_{k^\prime|k^\prime}^{i,(b_{k^\prime}(j))}$.
  \ENDFOR
  \STATE $\hat{x}_{k^\prime}^i = \frac{1}{P}\sum_{j=1}^P \widetilde{x}_{k^\prime}^j$.
  \ENDFOR
  \end{algorithmic}
  \label{alg1}
\end{algorithm}

\section{}\label{appendix_e}

In this appendix, we present the LP metric \cite[Proposition 2]{garcia2020metric} used to evaluate the multi-trajectory estimation performance in Section \ref{sec_performance_evaluation}.

For $1\leq p<\infty$, cut-off distance $c > 0$, track switching cost $\gamma > 0$ and a base metric $d_b(\cdot,\cdot)$ in the single object space $\mathfrak{X}$, the LP metric between sets $\mathbf{X}$ and $\mathbf{Y}$ of trajectories in time interval $1,\dots,K$ is given by
\begin{multline}
  \bar{d}_p^{c,\gamma}(\mathbf{X},\mathbf{Y}) = \underset{\substack{W^k\\k=1,\dots,K}}{\min} \left( \sum_{k=1}^K \text{tr}\left[\left(D^k_{\mathbf{X},\mathbf{Y}}\right)^{T}W^k\right]\right. \\
  + \left. \frac{\gamma^p}{2}\sum_{k=1}^{K-1}\sum_{i=1}^{|\mathbf{X}|}\sum_{j=1}^{|\mathbf{Y}|}\left| W^k(i,j) - W^{k+1}(i,j) \right|\right)^{\frac{1}{p}},
\end{multline}
where 
\begin{subequations}
  \begin{align}
    D^k_{\mathbf{X},\mathbf{Y}}(i,j) &= d\left(\mathbf{x}_i^k,\mathbf{y}_j^k\right)^p,\\
    d\left(\mathbf{x},\mathbf{y}\right) &= \begin{cases}
      \min(c,d_b(x,y)) & \mathbf{x} = \{x\}, \mathbf{y} = \{y\} \\
      0 & \mathbf{x} = \mathbf{y} = \emptyset \\
      \frac{c}{2^{1/p}} & \text{otherwise},
    \end{cases}\\
    &\sum_{i=1}^{|\mathbf{X}|+1}W^k(i,j) = 1, j = 1,\dots,|\mathbf{Y}|,\\
    &\sum_{j=1}^{|\mathbf{Y}|+1}W^k(i,j) = 1, i = 1,\dots,|\mathbf{X}|,\\
    &W^k(|\mathbf{X}|+1,|\mathbf{Y}|+1) = 0,\\
    &W^k(i,j) \geq 0, \forall i,j.
  \end{align}
\end{subequations}

The LP metric is computable in polynomial time, and it can be decomposed into costs for properly detected objects, missed and false objects, and track switches. We refer the readers to \cite{garcia2020metric} for implementation details.

}

\vfill

\end{document}